\documentclass{aa}  

\usepackage{graphicx}
\usepackage{gensymb}
\usepackage{lmodern}
\usepackage{txfonts}
\usepackage{caption}
\usepackage{lscape}
\usepackage{subfigure}
\usepackage{float}

\usepackage[]{hyperref}
\begin{document} 

   \title{FAUST X: Formaldehyde in the Protobinary System [BHB2007] 11: Small Scale Deuteration}

   \author{L. Evans
          \inst{1,2,3}
          \and
          C. Vastel\inst{1}
          \and
          F. Fontani\inst{2}
          \and
          J. E. Pineda\inst{4}
          \and
          I. Jim\'{e}nez-Serra\inst{5}
           \and
          F. Alves\inst{4}
           \and
          T. Sakai\inst{6}
           \and
          M. Bouvier\inst{7}
           \and
           P. Caselli\inst{4}
           \and
           C. Ceccarelli\inst{8}
           \and
           C. Chandler\inst{9}
           \and
           B. Svoboda\inst{9}
           \and
           L. Maud\inst{10}
           \and
           C. Codella\inst{2,8}
           \and
           N. Sakai\inst{11}
           \and
           R. Le Gal\inst{8}
           \and
           A. L\'{o}pez-Sepulcre\inst{8,12}
           \and
           G. Moellenbrock\inst{9}
           \and
           S. Yamamoto\inst{13,14}
 }

   \institute{IRAP, Universit\'{e} de Toulouse, CNRS, CNES, UPS, Toulouse, France\\
         \and
             INAF-Osservatorio Astrofisico di Arcetri, Largo E. Fermi 5, I-50125, Florence, Italy
         \and
            School of Physics and Astronomy, University of Leeds, Leeds LS2 9JT, UK
         \and
            Max-Planck-Institut f\"{u}r extraterrestrische Physik (MPE), Gie{\ss}enbachstr. 1, D-85741 Garching, Germany
        \and
            Centro de Astrobiologia (CSIC-INTA), Ctra. de Torrejon a Ajalvir, km 4, 28850, Torrejon de Ardoz, Spain
        \and
            Graduate School of Informatics and Engineering, The University of Electro-Communications, Chofu, Tokyo 182-8585, Japan
        \and
            Leiden Observatory, Leiden University, P.O. Box 9513, 2300 RA Leiden, The Netherlands
        \and
            Universit\'{e} Grenoble Alpes, CNRS, Institut de Plan\'{e}tologie et d’Astrophysique de Grenoble (IPAG), 38000 Grenoble, France
         \and
            National Radio Astronomy Observatory, PO Box O, Socorro, NM 87801, USA
        \and
            European Southern Observatory, Karl-Schwarzschild Str. 2, 85748 Garching bei M\"{u}nchen, Germany
        \and
            RIKEN Cluster for Pioneering Research, 2-1, Hirosawa, Wako-shi, Saitama 351-0198, Japan
        \and
            Institut de Radioastronomie Millim\'{e}trique (IRAM), 300 rue de la Piscine, 38406 Saint-Martin-d’H\`{e}res, France
        \and
            Department of Physics, The University of Tokyo, 7-3-1, Hongo, Bunkyo-ku, Tokyo 113-0033, Japan
        \and
            Research Center for the Early Universe, The University of Tokyo, 7-3-1, Hongo, Bunkyo-ku, Tokyo 113-0033, Japan
            \\
            \\
        \email{l.e.evans@leeds.ac.uk}
}

   \date{Received ; accepted }
  \abstract
    {Deuterium in H-bearing species is enhanced during the early stages of star formation, however, only a small number of high spatial resolution deuteration studies exist towards protostellar objects, leaving the small-scale structures of these objects unrevealed and understudied.}
   {We aim to constrain the deuterium fractionation ratios in a Class 0/I protostellar object in formaldehyde (H$_2$CO), which has abundant deuterated isotopologues in this environment.}
   {We observed the Class 0/I protobinary system [BHB2007] 11, whose emission components are embedded in circumstellar disks that have radii of 2 to 3 au, using the Atacama Large Millimeter Array (ALMA) within the context of the Large Program Fifty AU STudy of the chemistry in the disk/envelope system of Solar-like protostars (FAUST). The system is surrounded by a complex filamentary structure (the so-called streamers) connecting to the larger circumbinary disk. In this work we present the first study of formaldehyde D-fractionation towards this source with detections of H$_2$CO 3(0,3)-2(0,2), combined with HDCO 4(2,2)-3(2,1), HDCO 4(1,4)-3(1,3) and D$_2$CO 4(0,4)-3(0,3). These observations probe the structures of the protobinary system, enabling us to resolve multiple velocity components associated with the methanol hot spots also uncovered by FAUST data, as well as the colder external envelope. In addition, based on the kinematics seen in the observations of the H$_2$CO emission, we propose the presence of a second large scale outflow.}
   {From our ALMA observations, our results agree with current literature in only finding the deuterated species HDCO and D$_2$CO in the central regions of the core, while undeuterated H$_2$CO is found more ubiquitously. From the radiative transfer modelling, the column density of H$_2$CO ranges between (3-8)$\times$10$^{14}$ cm$^{-2}$ and that of HDCO ranges between (0.8-2.9)$\times$10$^{13}$ cm$^{-2}$. Column density for the single detected velocity component of D$_2$CO ranges between (2.6-4.3)$\times$10$^{12}$ cm$^{-2}$. This yields an average D/H ratio for formaldehyde in [BHB2007] 11 of 0.02$^{+0.02}_{-0.01}$ from HDCO
   . Following the results of kinematic modelling, the second large scale feature appears to be inconsistent with a streamer-like nature due to a flat and outflowing velocity relation; we thus tentatively conclude that the feature is an asymmetric molecular outflow launched by a wide-angle disk wind.}
{}

   \keywords{astrochemistry --
                radiative transfer --
                techniques: high angular resolution -- ISM: abundances -- ISM: molecules -- line: identification
               }
\titlerunning{FAUST IX: Formaldehyde in [BHB2007] 11: Deuteration and Kinematics}
\authorrunning{Evans et al.}

\maketitle

\section{Introduction}\label{Introduction}
 During the early prestellar core stage of star formation, the temperature is low (10 K) and the density is high - larger than $10^5$ cm$^{-3}$ in the central region (\citealt{caselli12}). This cold, dense environment encourages the formation of the deuterated species H$_2$D$^+$ (first detected by \citealt{stark}), which is an exothermic process (\citealt{gerlich}):
 \begin{equation}
\rm{H_3^+ + HD \rightleftharpoons H_2D^+ + H_2 + 232~K},
\label{eq.h2dp}
\end{equation}
while also giving rise to the depletion of CO due to carbon, oxygen and nitrogen-bearing species being frozen-out onto dust grains (\citealt{caselli99}, \citeyear{caselli02}). The rate of formation of H$_2$D$^+$ is slowed down through preferential reaction of the H$_3^+$ ion with CO to form HCO$^+$ instead of with HD to form H$_2$D$^+$:
\begin{equation}
\rm{H_3^+ + CO \rightarrow HCO^+ + H_2}
\label{eq.co1}
\end{equation}
and thus the absence of CO in these environments also favours the production of the deuterated species. In fact, we can even go beyond this with the production of D$_2$H$^+$ (first detected by \citealt{Vastel04}) and D$_3^+$.
\\
\\
The protostellar phase is initiated due to cloud collapse as a result of gravity overcoming pressure. This generally occurs when the cloud mass reaches a limit known as Jeans mass, although this value can change during the collapse process, leading to the formation of multiple cores. During the evolutionary process, the temperature increases up to approximately 100 K; CO is released back into the gas phase from the dust grains when the temperature surpasses the critical value of approximately 25 K. Therefore, as the embedded protostar evolves and more species are injected into the gas phase via sublimation from the grain surface, the destruction of H$_2$D$^+$, D$_2$H$^+$ and D$_3^+$ does occur through gas phase reactions with neutrals. In particular, reactions with CO and N$_2$ lead to an increase in DCO$^+$ and N$_2$D$^+$, respectively; all of these deuterated ions eventually recombine in the gas phase to form, alongside other species, D atoms, which lead to the formation of neutral deuterated species and an increase in the D/H ratio in the gas phase. Recombination also takes place on the icy surfaces of dust grains allowing deuteration to also occur in this phase. 
Formaldehyde (H$_2$CO) is already present in the core by this stage, therefore, in theory, deuterium in its increased abundance will form deuterated formaldehyde.
\\
\\
Multiple observations have confirmed the presence of both HDCO and D$_2$CO in protostellar cores (e.g. \citealt{loinard}, \citealt{parise_06}, \citealt{persson}, \citealt{sahu}). The D/H ratio of protostellar cores varies by species but molecular D/H enhancements of up to 13 orders of magnitude compared to the cosmic value of $10^{-5}$ have been observed (\citealt{loinard}, \citealt{parise_04}, \citealt{ceccarelli}), including D/H ratios of up to 30\% in D$_2$CO (\citealt{loinard}). More recently, \citet{persson} found the formaldehyde D/H ratio in the low mass protostar IRAS 16293-2422 B to be 0.03 for HDCO and 0.08 for D$_2$CO.
\\
\\
The exact deuteration pathway for formaldehyde is not conclusively known. Two scenarios have been presented because undeuterated H$_2$CO can be formed both in the gas phase (\citealt{parise_09}, \citealt{yamamoto}) and on grain surfaces (\citealt{watanabe}, \citealt{roberts}, \citealt{bergman}). Therefore, deuteration pathways in both phases have been explored. The grain surface pathway, proposed by \citet{rodgers}, suggests that (deuterated) formaldehyde is formed in intermediate steps of the (deuterated) CH$_3$OH formation pathway involving (D-addition or) hydrogenation. Meanwhile, the gas phase pathway has been proposed to involve formaldehyde deuteration either directly from reaction with H$_2$D$^+$ or via the deuteration of CH$_3^+$ (\citealt{roberts}, \citealt{roueff}), which showed consistency with subsequent modelling predictions (\citealt{bergman}). However, observational analyses (\citealt{taquet}, \citealt{ceccarelli14}, \citealt{fontani}, \citealt{persson}, \citealt{manigand}, \citealt{zahorecz}), as well as chemical modelling (\citealt{rodgers}), favour the grain surface deuteration pathway as the most likely for formaldehyde.
\\
\\
Despite all of this previous research, our knowledge of the fractionation processes in H$_2$CO is not yet clear especially at small linear scales, since previous results have been mostly obtained through single-dish observations and represent, therefore, the average values of regions known to have density and temperature variations at small scales. Interferometric observations are therefore required to probe deeper into protostellar objects and reveal the chemistry within. In particular, the object of this study, [BHB2007] 11, is very understudied compared to other similar sources, for example, IRAS 16293-2422 B (\citealt{ceccarelli01}, \citealt{bottinelli}, \citealt{crimier}, \citealt{persson}). With this in mind, [BHB2007] 11 was chosen as one of 13 Class 0 and Class I sources to be observed with the Atacama Millimeter/Submillimeter Array (ALMA) within the context of the Large Program Fifty AU STudy of the chemistry in the disk/envelope system of Solar-like protostars (FAUST\footnote{http://stars.riken.jp/faust/fausthome.html}; PI: Satoshi Yamamoto; see also \citealt{codella}), with setups chosen, in part, with the goal of observing multiple deuterated species. Additionally, observing the core with multiple configurations enables both large- and small-scale analysis of our source. These observations will provide a new perspective to the chemical analysis of a protostellar core. In addition, the present work represents the first deuteration study towards [BHB2007] 11.
\subsection{Source Background}\label{Source_Background}
[BHB2007] 11 is a Class 0/I protostellar core (\citealt{brooke},  \citealt{forbrich},  \citealt{sandell}) located 163 $\pm$ 5 pc away (\citealt{dzib}) in the Barnard 59 (B59) core. The source is among at least 20 low mass Young Stellar Objects (YSOs) forming a protocluster within B59. While the members of this protocluster have been shown to exhibit diversity in evolutionary stage (\citealt{onishi}), the YSO [BHB2007] 11 is believed to be the youngest among the cluster at approximately 0.1-0.2 Myr old (\citealt{brooke},  \citealt{riaz}). This object has a bolometric temperature of 60-70 K, luminosity of 2.2-4.5 L$_\odot$ and systemic velocity of 3.6 km~s$^{-1}$ (\citealt{onishi}, \citealt{brooke}, \citealt{forbrich}, \citealt{sandell}). Previous observations have revealed [BHB2007] 11 as a protobinary system with two central objects (hereafter denoted A ($\alpha$(2000)=17h11m23.1058s, $\delta$(2000)=-27\degree24'32.828") and B ($\alpha$(2000)=17h11m23.1015s, $\delta$(2000)=-27\degree24'33.987"), respectively, located approximately 28 au apart (in projection) with A to the north of B (\citealt{alves_19}). Source B has been shown to exhibit preferential accretion from the circumbinary disk, traced by infalling gas within filaments (\citealt{alves_19}), 
while Source A is the more massive of the two protostars. Preferential accretion into the less massive companion of a protobinary system has been predicted by simulations (\citealt{bate_97},  \citealt{bate_02},  \citealt{matsumoto}). 
\\
\\
In addition, recent observations have shown the presence of narrow structures from beyond the core down to disk scales in free-fall, which are named streamers (see \citealt{pineda22} for a review), while observations of CO, C$^{18}$O and H$_2$CO have revealed that this core features a symmetrical bipolar outflow. It has also been determined that this outflow is launched from outside the disk edge (at about 100 au; \citealt{Alves17}), which is unusual for this type of object. Using H$_2$CO observations, \citet{Alves17} also detected the existence of a centrifugal barrier marking a sharp change in kinematics within the spiral structure which itself is revealed in continuum (\citealt{Alves17}). Position-velocity (PV) diagrams towards [BHB2007] 11 have varying profiles depending on the species; while higher energy transitions ($E_{\rm{u}}~\sim$ 70 K) of H$_2$CO showed a smooth Keplerian PV plot profile (\citealt{alves_19}), CH$_3$OH and lower energy transitions of H$_2$CO ($E_{\rm{u}}~\sim$ 20 K) showed a more fragmented profile consistent with multiple velocity components (\citealt{Alves17}, \citealt{vastel}). This CH$_3$OH emission has been detected in three identified positions within 30 au of A and B in the circumstellar region (in contrast to the more ubiquitously detected H$_2$CO) and is believed to originate from the sublimation of dust grains due to non-thermal processes, such as the existence of a shocked region. These shocks would most likely be caused by the interaction between quiescent gas and previously detected streamers connecting the circumbinary disk to the circumstellar region (\citealt{alves_19}, \citealt{vastel}).  
\\
\\
The structure of this paper is set out as follows: Sect. \ref{Observations} describes the observations, Sect. \ref{Methods} describes the emission morphology, Sect. \ref{radtrans} discusses the radiative transfer analysis and Sect. \ref{Conclusions} contains the conclusions.
\begin{table*}[!h]
    \footnotesize
    \centering
    \caption{ALMA Observational parameters for [BHB2007] 11 in setups 1 and 2.}
    \begin{tabular}{c c c c c c c
    }
    \hline
    \hline
         & & Setup 1 & & & Setup 2 & 
         \\
         \hline
         \hline
        & & & & & & 
        \\
        ALMA Band & & 6 & & & 6 & 
        \\
        & & & & & & 
        \\
        \hline
        & & & & & & 
        \\
        Frequency & & 214.0-219.0 & & & 242.5-247.5 & 
        \\
        range [GHz] & & 229.0-234.0 & & & 257.5-262.5 & 
        \\
        & & & & & & 
        \\
        \hline
        \hline
        Configuration & C43-4 & C43-1 & 7M & C43-4 & C43-1 & 7M 
        \\ 
        \hline
        \hline
        & & & & & & 
        \\
        Diameter [m] & 12 & 12 & 7 & 12 & 12 & 7 
        \\
        & & & & & & 
        \\
        \hline
        & & & & & & 
        \\
        Baseline [m] & 15-780 & 15-160 & 0-33 & 15-780 & 15-160 & 0-33 
        \\
        & & & & & &
        \\
        \hline
        & & & & & & 
        \\
        Beamsize ["] & 0.35 $\times$ 0.33 & 1.73 $\times$ 1.25 & 7.71 $\times$ 3.94 & 0.35 $\times$ 0.33 & 1.73 $\times$ 1.25 & 7.71 $\times$ 3.94 
        \\
        $\theta_{\rm max}\times\theta_{\rm min}$ & & & & & &\\
        & & & & & &
        \\
        \hline
        & & & & & &
        \\
        Maximum & 11.4 & 11.5 & 19.2 & 11.4 & 11.5 & 19.2 
        \\
        Recoverable Scale ["] & & & & & &
        \\
        & & & & & &
        \\
        \hline
        & & & & & &
        \\
        Water vapour [mm] & 1.4-2.4 & 1.4-2.4 & 1.14-1.30 & 1.6-2.0 & 1.6-2.0 & 1.06-1.31
        \\
        & & & & & & 
        \\
        \hline
    \end{tabular}
    \label{table_obs}
\end{table*}
\normalsize
\section{Observations}\label{Observations}
Within the context of the Large Program FAUST, the Class 0/I protostellar core [BHB2007] 11 was observed in the ALMA large project 2018.1.01205.L between 2018 and 2020 with spectral resolution 0.14 MHz (equivalent to a velocity resolution of 0.2 km~s$^{-1}$) in setup 1 and 0.12 MHz (equivalent to a velocity resolution of 0.2 km~s$^{-1}$) in setup 2, along with angular resolution 0.25" (equivalent to a spatial resolution of $\sim$ 40 au) across the setups. This source was observed with the antennas configured in three different ways: two configurations with the main array (12 m)
and one configuration with the Atacama Compact Array (ACA, 7 m antennas)
. In this paper, we present observations derived from the combination of these 12 m and 7 m configurations using the 'mosaic' gridder in \textit{tclean} within CASA. Our combination process involves identifying line-free continuum channels common to all datasets and configurations by hand, then using self-calibration of the resulting continuum to align both phases and amplitudes to take out position and amplitude calibration offsets between execution blocks. During this process, we take care to make sure that the resulting continuum model is as complete as possible to avoid introducing any scaling offsets that can suppress the overall amplitude scale. The final robust weighting used for the continuum is one chosen to deliver 50 au resolution (as required for the FAUST project) but natural weighting is used during the self-calibration to ensure all the emission is recovered by the sky model during the imaging, resulting in a beamsize of 0.43" $\times$ 0.41". The FAUST program provides observations in Bands 3 and 6 divided in three spectral setups, for a total of 13 spectral windows (spws). In this paper we analyse lines from Band 6 only, since the frequency range covers the formaldehyde isotopologues that we are targeting. Thus in Table \ref{table_obs} we list the frequency ranges only of the windows included in these setups. Table \ref{table_obs} also contains useful observational information, such as the angular resolution and the maximum recoverable scale achieved in each configuration and setup, as well as water vapour conditions. The system temperature ranged from 50-160 K. 
For 12 m observations in setups 1 and 2 the quasars J1427-4206 and J2056-4714 were used as bandpass calibrators, while J1700-2610 was used to calibrate phase and flux. The 1.3 mm continuum was observed with spectral resolution 0.98 MHz (setup 1) and 1.13 MHz (setup 2). 
\begin{table}[!h]
    \centering
    \caption{Spectroscopic parameters for all transitions in the sample, along with \textit{rms} associated with each corresponding spw.}
    \tiny
    \begin{tabular}{c c c c c c}
    \hline
         Species & Frequency & Transition & $A_{\rm ij}$ & $E_{\rm u}$ & \textit{rms} \\
         & (MHz) & & (s$^{-1}$) & (K) & (K) \\
         \hline
         \hline
         H$_2$CO & 218222.192 & 3(0,3)-2(0,2) & 2.82$\times$10$^{-4}$ & 20.96 & 0.15 \\
         \\
         HDCO & 246924.600 & 4(1,4)-3(1,3) & 3.96$\times$10$^{-4}$ & 37.60 & 0.021\\
         \\
         HDCO & 259034.910 & 4(2,2)-3(2,1) & 3.66$\times$10$^{-4}$ & 62.87 & 0.072 \\
         \\
         D$_2$CO & 231410.234 & 4(0,4)-3(0,3) & 3.47$\times$10$^{-4}$ & 27.88 & 0.15 \\
         \hline
    \end{tabular}
    \label{table_transitions}
\end{table}
\noindent Table \ref{table_transitions} shows the spectroscopic parameters of all of the formaldehyde transitions, including singly (HDCO) and doubly (D$_2$CO) deuterated transitions, that were identified within the FAUST dataset using CASSIS\footnote{Based on analysis carried out with the CASSIS software (\citealt{Vastel2015}) and JPL (http://spec.jpl.nasa.gov/) and CDMS (https://cdms.astro.uni-koeln.de/classic/) molecular databases. CASSIS has been developed by IRAP-UPS/CNRS (http://cassis.irap.omp.eu)} software.
\\
\\
The data were reduced, calibrated and cleaned using the Common Astronomy Software Application (CASA\footnote{CASA is developed by an international consortium of scientists based at the National Radio Astronomical Observatory (NRAO), the European Southern Observatory (ESO), the National Astronomical Observatory of Japan (NAOJ), the Academia Sinica Institute of Astronomy and Astrophysics (ASIAA), the CSIRO division for Astronomy and Space Science (CASS) and the Netherlands Institute for Radio Astronomy (ASTRON) under the guidance of NRAO.}, {bf CJC: version 5.6.1-8}) utilising a pipeline developed by the FAUST data reduction team (see \citealt{vastel} for more details), with a Briggs robust uv-weighting of 0.5.
\section{Emission Morphology}\label{Methods}
The clean maps were 
analysed with the GILDAS\footnote{The GILDAS software is developed at the IRAM and the Observatoire de Grenoble and is available at http://www.iram.fr/IRAMFR/GILDAS} software. 
The line channels were identified and used to create moment 0 maps for each of the transitions. In order to investigate both large- and small-scale structure, we combine observations performed using both 12 m and 7 m ALMA configurations in all moment 0 maps shown in this section. For the analysis of all lines in our sample, the 4$\sigma$ contour (the minimum to define a detection) of the map of the HDCO transition at 259 GHz was used to define a polygon for extraction. This transition was chosen to estimate the D/H ratio because it is detected at a signal-to-noise ratio (s/n) of 8$\sigma$ and because the HDCO transition at 246 GHz shows possible signs of blending with the CH$_3$OCHO (19-18) transition.
\\
\\
The spectra of the four transitions were extracted in flux density units (S$_\nu$) within the aforementioned polygon; these were converted to synthesised beam temperature units (K) in the GILDAS package CLASS using conversion factors calculated using Eq. \ref{eq.conversion}, which was taken from the ALMA handbook (\citealt{warmels}):
\begin{gather}
    \Bigg(\frac{T}{1 K}\Bigg)=\Bigg(\frac{S_\nu}{1 Jy}\Bigg)\Bigg[13.6\Bigg(\frac{300 GHz}{\nu}\Bigg)^2\Bigg(\frac{1''}{\theta_{max}}\Bigg)\Bigg(\frac{1''}{\theta_{min}}\Bigg)\Bigg],
    \label{eq.conversion}
\end{gather}
where $\theta_{\rm min}$ and $\theta_{\rm max}$ represent the half power beam widths along the minor and major axes, respectively.
\subsection{H$_2$CO 3(0,3)-2(0,2)}\label{H2CO}
\begin{figure*}[!h]
    \centering
    \begin{subfigure}{}
    \includegraphics[width=230pt]{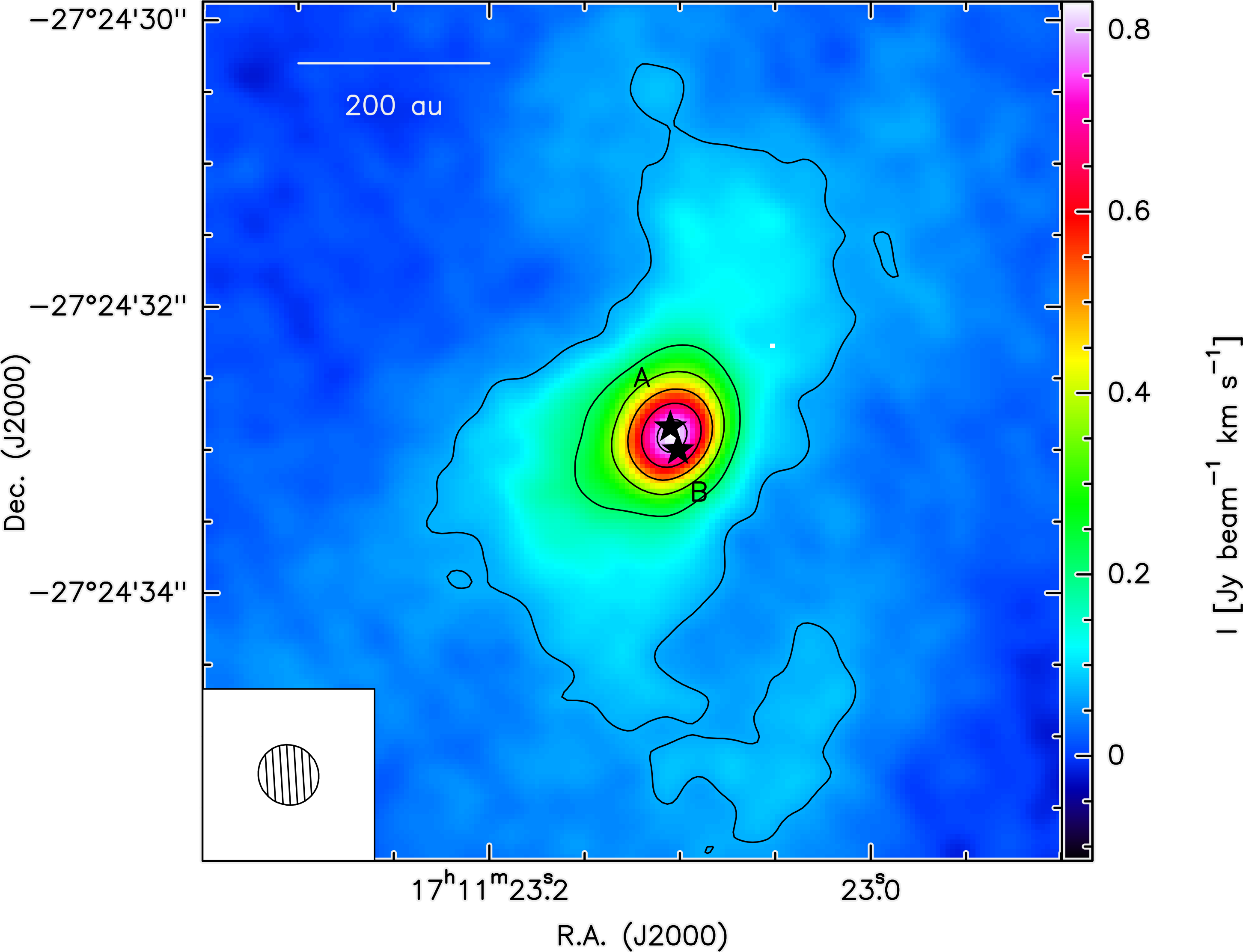}
    \end{subfigure}
    \begin{subfigure}{}
    \includegraphics[width=230pt]{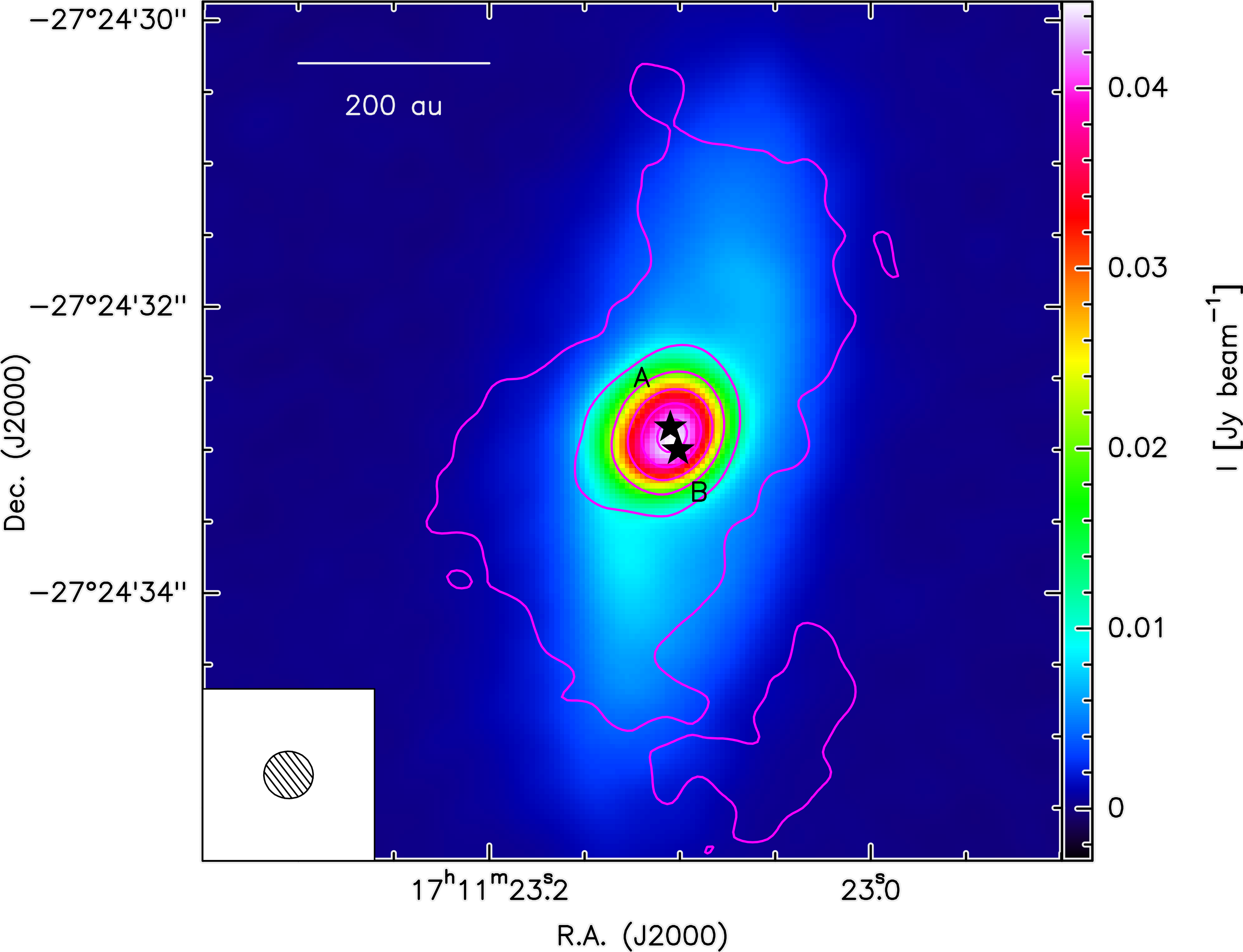}
    \end{subfigure}
    \caption{Left panel: H$_2$CO 3(0,3)-2(0,2) moment 0 map integrated between -5 and 15 km~s$^{-1}$. This velocity range was chosen as this represents the full range of emission detected at or above 4$\sigma$. Right panel: 216 GHz continuum emission (colour scale) taken from the FAUST dataset with H$_2$CO 3(0,3)-2(0,2) overlaid in pink contours. In both images, contours start at 4$\sigma$ and end at 44$\sigma$ in 8$\sigma$ increments ($\sigma$=17.6 mJy~beam$^{-1}$~km~s$^{-1}$), the positions of protostars A and B are represented by the black stars and the ALMA synthesised beam is shown in the lower left corner.}
    \label{fig.H2CO_cont}
\end{figure*}
\begin{figure*}[!h]
    \centering
    \begin{subfigure}{}
    \includegraphics[width=250pt]{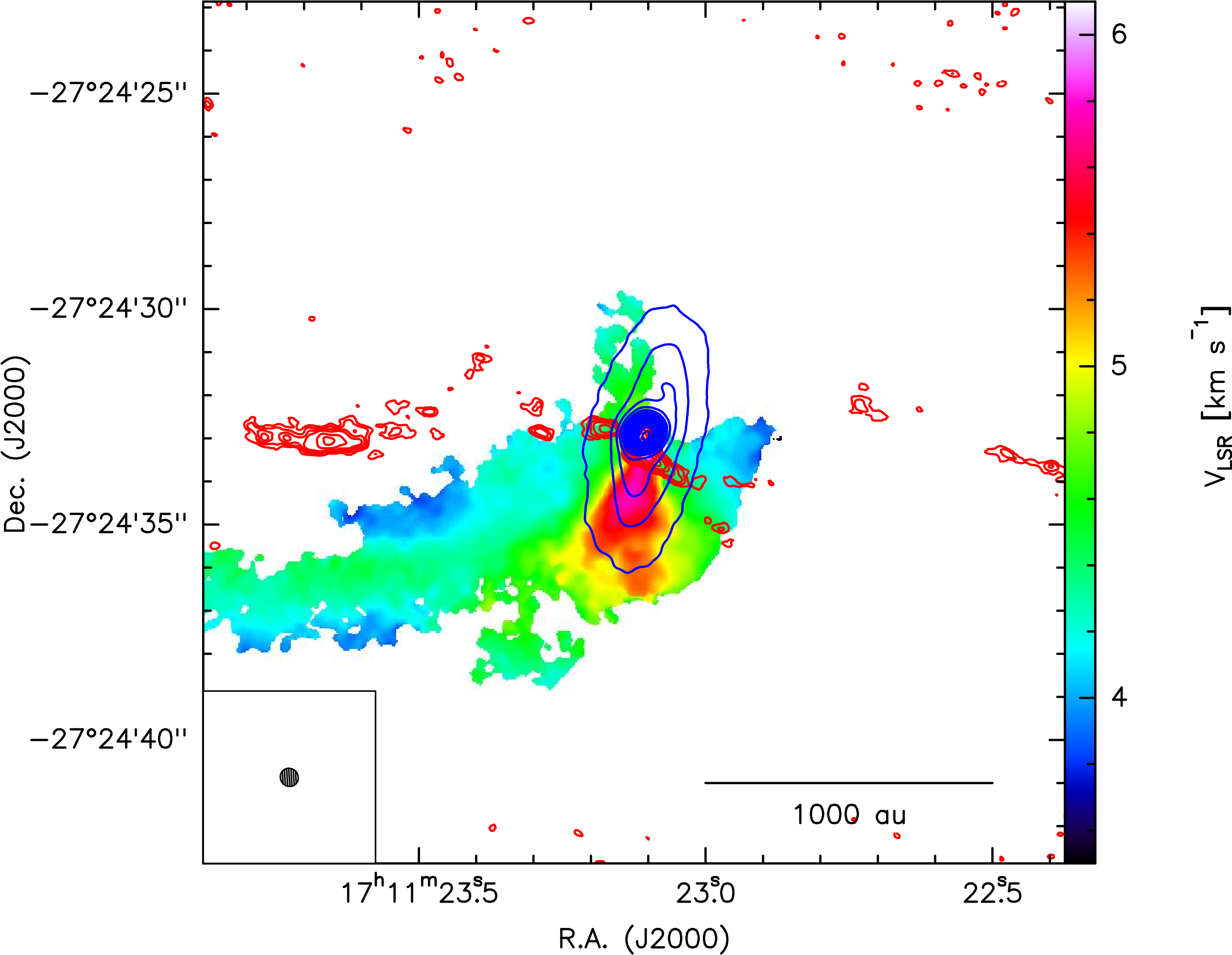}    
    \end{subfigure}
    \begin{subfigure}{}
    \includegraphics[width=250pt]{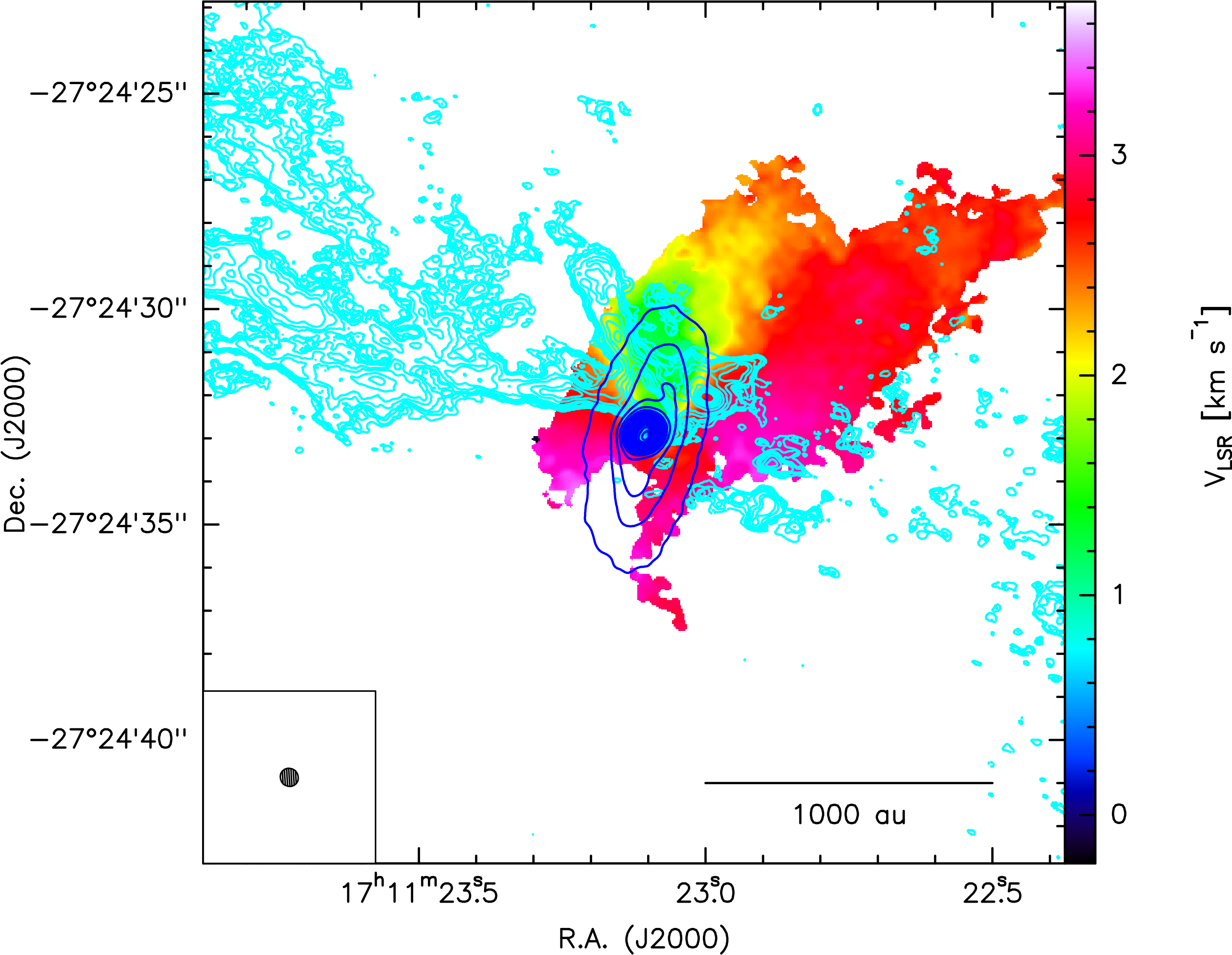}
    \end{subfigure}
    \caption{Colour scale: H$_2$CO 3(0,3)-2(0,2) centroid velocity map from pixel-by-pixel Gaussian analysis showing pixels with intensity above 5$\sigma$ in the redshifted (left panel; integrated between 3.5 and 6 km~s$^{-1}$) and blueshifted (right panel; integrated between 0 and 3.5 km~s$^{-1}$) regimes; the systemic velocity is $\sim$3.6 km~s$^{-1}$. Blue contours: 1.3 mm continuum contours starting at 4$\sigma$ and ending at 300$\sigma$ in intervals of 15$\sigma$ ($\sigma$=0.19 mJy~beam$^{-1}$) indicating disk emission. Red (left panel) and cyan (right panel) contours: redshifted and blueshifted (respectively) CO 2-1 emission from \citet{Alves17} detailing outflow emission seen at 8.67 km~s$^{-1}$ and -0.78 km~s$^{-1}$, respectively, representing the channels with the most extended outflow CO 2-1 emission. Contours are at 4, 6, 8, 10, 15, 20, 30, 40, 50, 60, 70, 80 and 90 times $\sigma$, which is 2.8 mJy~beam$^{-1}$, in order to match the CO 2-1 observations, which have a beamsize of 0.2" $\times$ 0.25". The ALMA synthesised beam for the H$_2$CO observations, with a beamsize of 0.43" $\times$ 0.41" is shown in the lower left corner.}
    \label{fig.streamcont}
\end{figure*}
\noindent The emission morphology of H$_2$CO 3(0,3)-2(0,2) can be seen in the moment 0 map shown in the left panel of Fig. \ref{fig.H2CO_cont}. Kinematic analysis performed in previous studies (\citealt{Alves17}) and also later in this paper shows that this is a disk/outflow system. The axis of the extension seems to be aligned with that seen in the continuum, which has a position angle of 138$\degree$ (\citealt{Alves17}); this is illustrated in the right panel of Fig. \ref{fig.H2CO_cont}, which overlays the H$_2$CO 3(0,3)-2(0,2) emission morphology onto the 1.3 mm dust emission, which are clearly coincident. 
\begin{figure*}[!h]
    \centering
    \captionsetup[subfigure]{labelformat=empty}
    \begin{subfigure}{}
    \includegraphics[width=230pt]{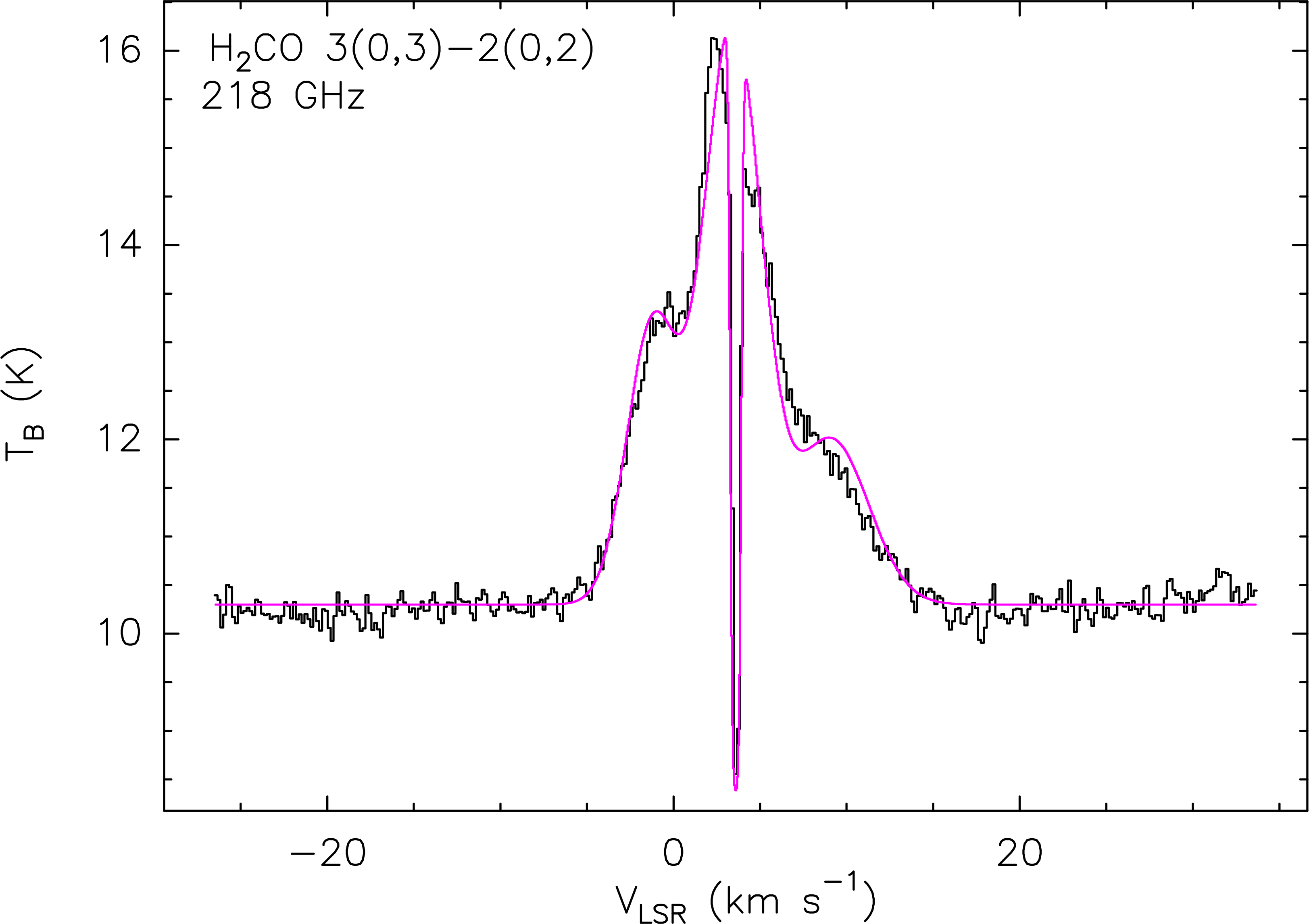}
    \end{subfigure}
    \begin{subfigure}{}
    \includegraphics[width=235pt]{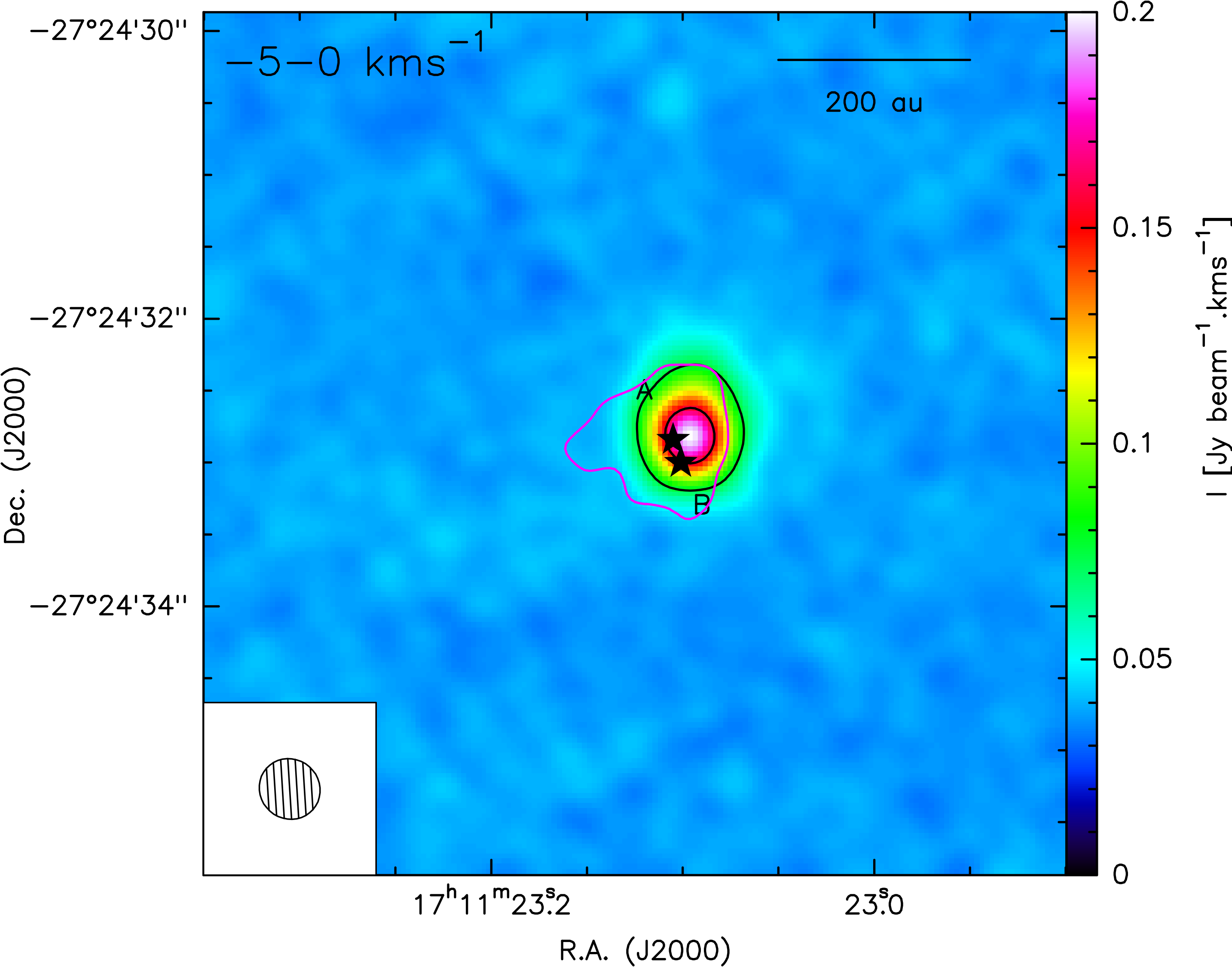}
    \end{subfigure}
    \begin{subfigure}{}
    \includegraphics[width=235pt]{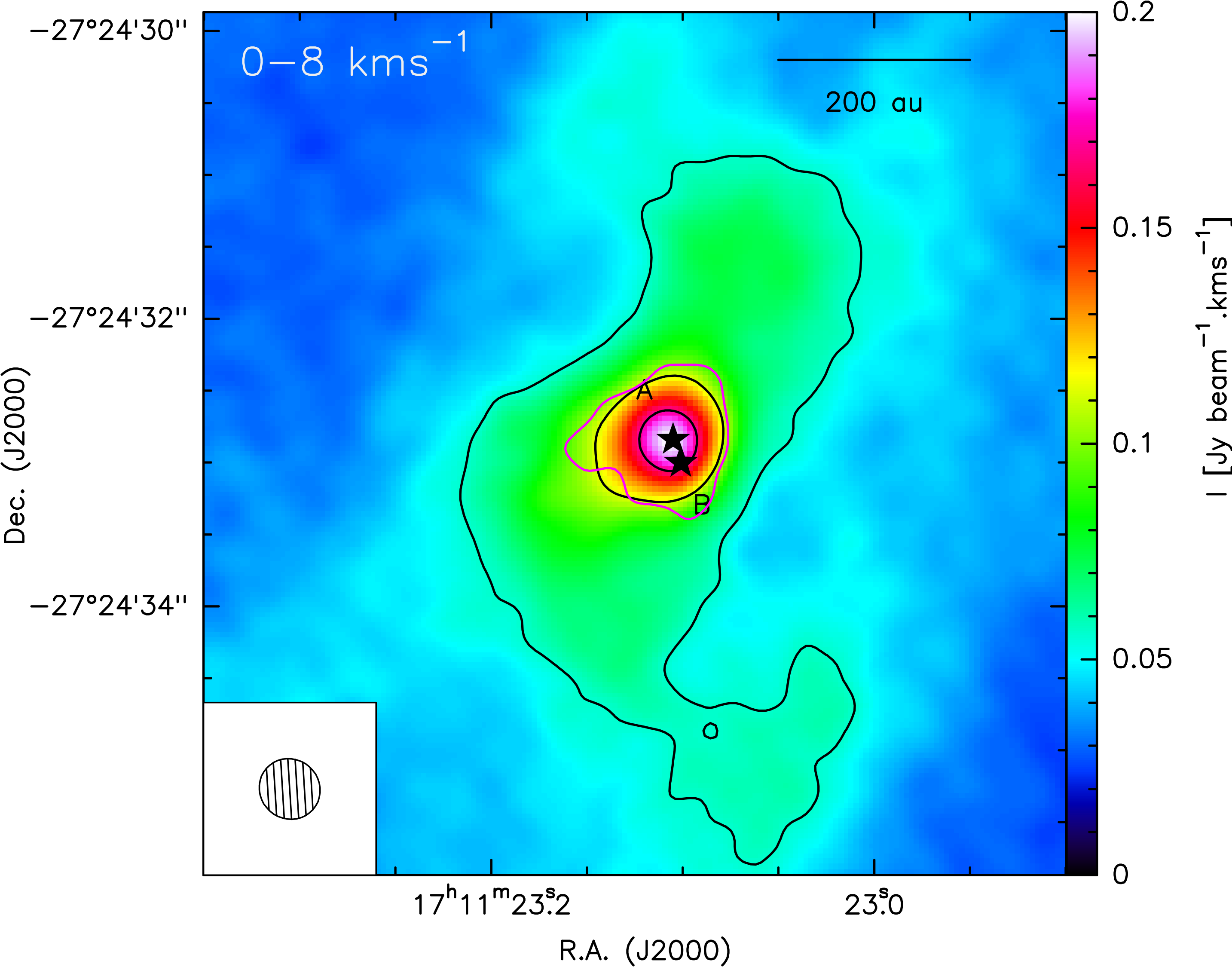}
    \end{subfigure}
    \begin{subfigure}{}
    \includegraphics[width=235pt]{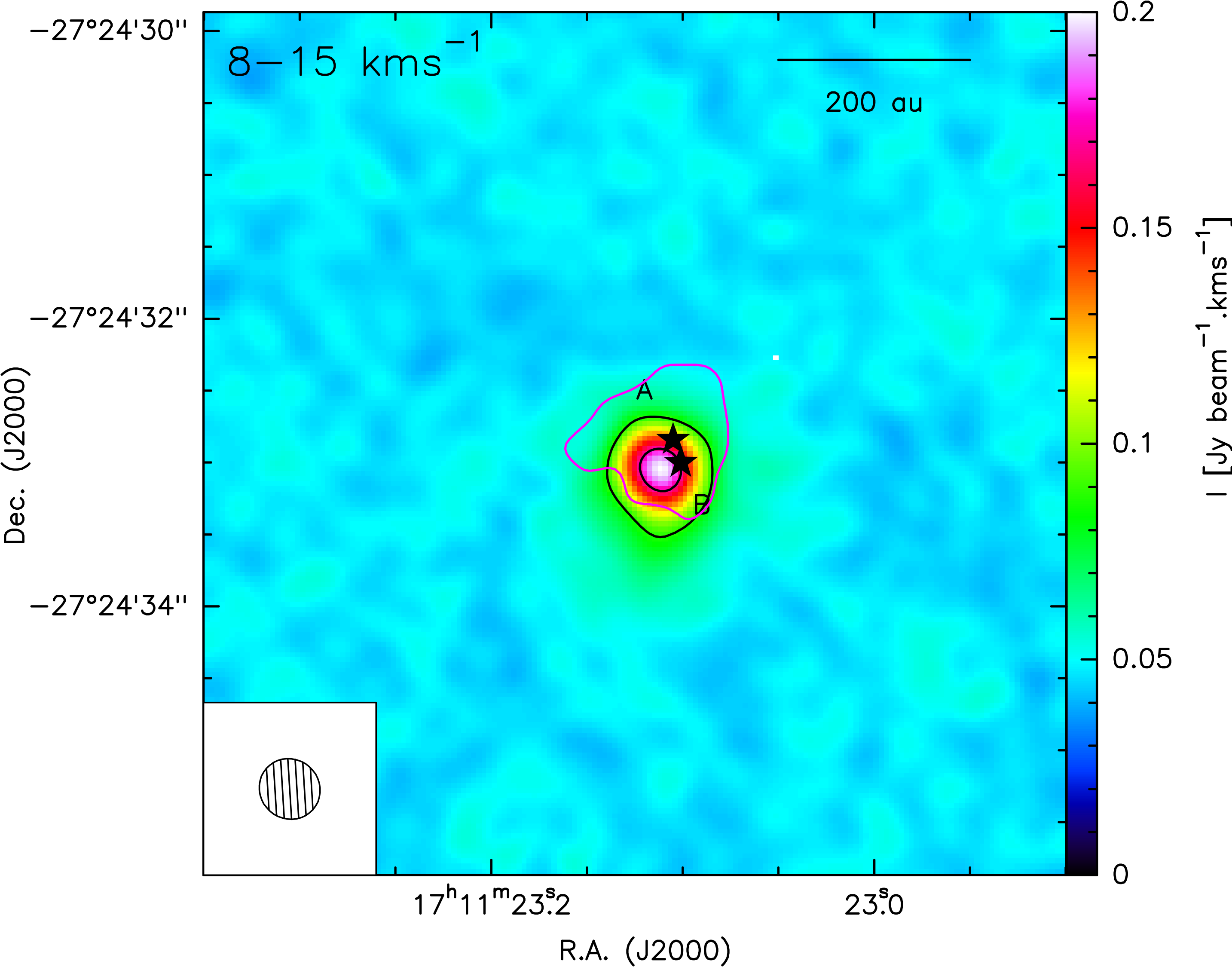}
    \end{subfigure}
    \caption{Upper left: H$_2$CO 3(0,3)-2(0,2) spectrum (black) with velocity resolution 0.17 km~s$^{-1}$ covering the full velocity range of -5 to 15 km~s$^{-1}$, extracted from the region where HDCO 4(2,2)-3(2,1) emission is above 4$\sigma$, overlaid with the best fit (pink) obtained using four Gaussian fits with FWHM, v$_{\rm{LSR}}$ and $T_{\rm{ex}}$ constrained using values obtained from fitting the HDCO transitions. A continuum value of 10.3 K has been added to this spectrum in order to model the absorption feature, located at the systemic velocity of 3.6 km~s$^{-1}$. Upper right, lower left, lower right: H$_2$CO 3(0,3)-2(0,2) moment 0 map integrated between -5 and 0 km~s$^{-1}$, between 0 and 8 km~s$^{-1}$ and between 8 and 15 km~s$^{-1}$ (respectively), in order to individually show the velocity components of emission shown in the upper left panel. Contours begin at 4$\sigma$ and are in increments of 8$\sigma$ ($\sigma$ = 17.6 mJy~beam$^{-1}$~km~s$^{-1}$). The pink contour illustrates the area within which the spectrum in the upper left panel was extracted, defined as the 4$\sigma$ contour of the HDCO 4(2,2)-3(2,1) moment 0 map (see Fig. \ref{fig.HDCO_mom1}). The ALMA synthesised beam is in the lower left corner of all three maps.}
    \label{H2CO_3mom0}
\end{figure*}
\\
\\
Fig. \ref{fig.streamcont} shows the centroid velocity maps obtained from a pixel-by-pixel Gaussian analysis using the \textit{pyspeckit} package (\citealt{ginsburg}) overlaid with contours from 1.3 mm emission from the FAUST data. For this Gaussian fitting we include pixels with velocities between 3.5 and 6 km~s$^{-1}$ (redshifted lobe; left panel of Fig. \ref{fig.streamcont}) and between 0 and 3.5 km~s$^{-1}$ (blueshifted lobe; right panel of Fig. \ref{fig.streamcont}) and a minimum intensity of 5$\sigma$. The redshifted and blueshifted high velocity extensions of the disk emission are clearly visible to the NW and SE, along with lower velocity extensions seen closer to the E/W axis. The high velocity emission coincides with the morphology of the continuum emission and thus appears to be originating from the disk, while the lower velocity emission is believed to be tracing dynamic material extending beyond the circumbinary disk, potentially originating from incoming streamer material or an outgoing disk wind. Kinematics of the formaldehyde emission suggest the presence of a second outflow (see Appendix for details). 5$\sigma$ was chosen as the threshold for the centroid velocity maps because, as explored in the Appendix, this is the optimal s/n to model dynamic emission without dilution from surrounding emission. To obtain the maps in Fig. \ref{fig.streamcont}, the velocity limits for integration were chosen as 0 and 6 km~s$^{-1}$ because the extended structure is not seen outside these velocities and therefore this velocity range is optimal for detecting and discerning both compact (disk) and extended (incoming streamer or outgoing disk wind) emission simultaneously.
It should be noted that, despite observing the same transition, the higher spatial resolution H$_2$CO observations published in \citet{Alves17} only appear to trace the disk emission aligned with the continuum emission and not the circumbinary large scale flow. The accretion filaments surrounding the binary system itself are a distinct feature at a much smaller scale. The reason for this is that our shortest baseline and overall uv coverage are more sensitive to extended emission than those in \citet{Alves17}. Notably, as illustrated in Fig. \ref{fig.streamcont}, the velocity structure does not show alignment with the outflow seen in CO 2-1 (\citealt{Alves17}) either in terms of morphology or velocity. The interferometric nature of these observations does lead to the possibility that some low velocity CO emission may be filtered out, however, the region affected by this is spatially much more extended than the region where deuterated species are detected (as will be described in subsequent sections) and thus the main conclusions of this work are unaffected. The features appear almost perpendicular to one another and at differing velocities, with the feature seen in H$_2$CO appearing closer to the systemic velocity of 3.6 km~s$^{-1}$ while the feature traced by CO peaks at more extreme redshifted and blueshifted velocities. It should, however, be noted that the beamsize of the observations presented in \citet{Alves17} is 0.20" $\times$ 0.25", a mismatch with the present FAUST data, which has a beamsize of 0.43" $\times$ 0.41".
\\
\\
\noindent As can be seen in Fig. \ref{H2CO_3mom0}, the H$_2$CO spectrum shows three clear emission components at $\sim$ -5 to 0 km~s$^{-1}$, $\sim$ 1 to 8 km~s$^{-1}$ and $\sim$ 8 to 15 km~s$^{-1}$. In order to further illustrate this and show the origin of each component, Fig. \ref{H2CO_3mom0} also shows individual moment 0 maps for each of the three emission components (integrated between -5 and 0 km~s$^{-1}$, 1 and 8 km~s$^{-1}$ and between 8 and 15 km~s$^{-1}$). The three emission components are broadly consistent with those seen in CH$_3$OH towards this object, which were seen at approximately -2, 2.8 and 9.9 km~s$^{-1}$ (\citealt{vastel}). Evidently, as seen in CH$_3$OH, the two H$_2$CO emission components at more extreme velocities are originating from the rotating disk as emission from material moving towards Source B (previously observed at high velocities by \citealt{alves_19}), while the central component originates from Source A and the surrounding envelope. Sect. \ref{origin} will explore and compare the origins of CH$_3$OH and H$_2$CO emission in this source in more detail. Additionally, H$_2$CO exhibits an absorption feature ($\sim$ 2 to 5 km~s$^{-1}$) due to absorption in the cold envelope of emission originating from warm inner regions, which, according to pixel-by-pixel analysis, is seen in a circular region that extends to approximately 500 au from the centre and is centred on the dust peak. Note that in order to model this absorption feature, a continuum value of 10.3 K has been added to the spectrum shown in the upper left panel of Fig. \ref{H2CO_3mom0}. This value has been obtained by measuring the flux of the 216 GHz continuum emission.
\subsection{HDCO 4(2,2)-3(2,1)}
\begin{figure*}[!h]
    \centering
    \begin{subfigure}{}   
    \includegraphics[width=230pt]{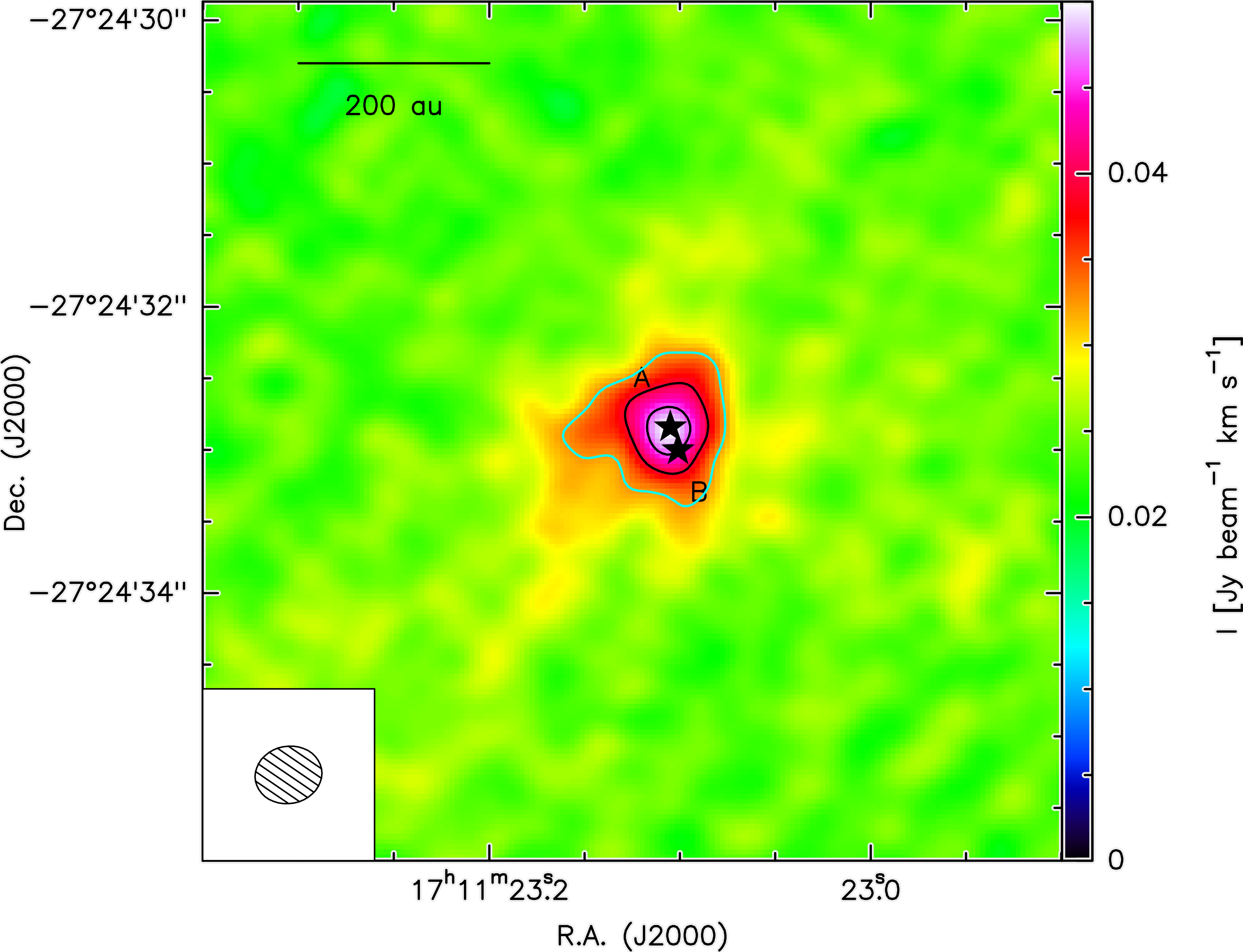}
    \end{subfigure}
    \begin{subfigure}{}
    \centering
    \includegraphics[width=230pt]{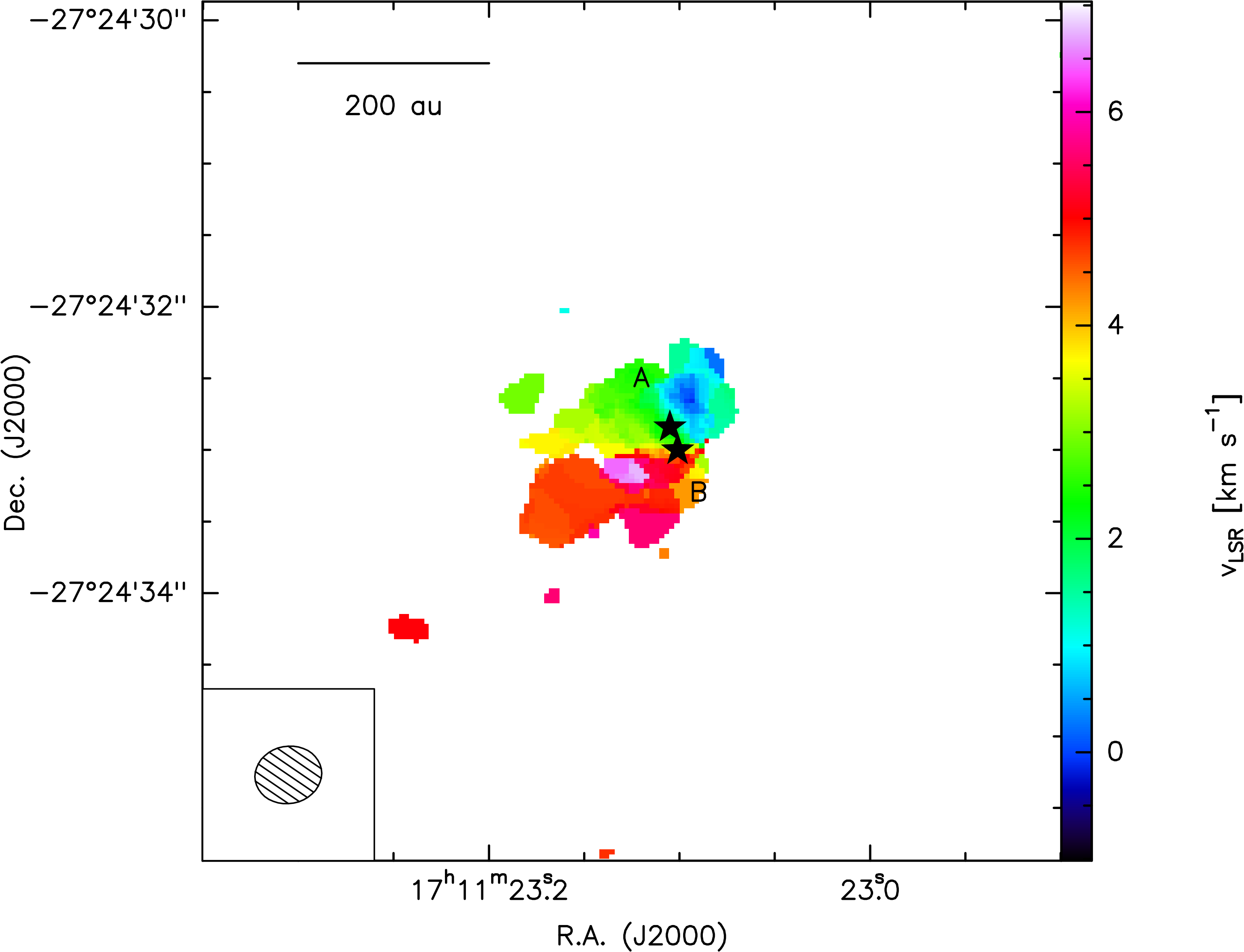}
    \caption{Left panel: HDCO 4(2,2)-3(2,1) moment 0 map integrated between -1 and 7 km~s$^{-1}$ that peaks at the position of the dust emission. Observations performed using both 12 m and 7 m ALMA configurations are combined in this map in order to investigate both large- and small-scale structure. Contours start at 4$\sigma$ and end at 10$\sigma$ in 3$\sigma$ increments ($\sigma$=4.1 mJy~beam$^{-1}$~km~s$^{-1}$). The 4$\sigma$ contour is highlighted in cyan as this is the contour that served as the extraction polygon for all spectra shown in this work. Right panel: Moment 1 map of HDCO 4(2,2)-3(2,1) integrated between -1 and 7 km~s$^{-1}$, in order to directly compare with the moment 0 map and to show velocity structure of the disk. In both maps, the locations of the protostars A and B are shown by black stars. The ALMA synthesised beam is in the lower left corner.}
    \label{fig.HDCO_mom1}
    \end{subfigure}    
\end{figure*}
\noindent The emission morphology for HDCO 4(2,2)-3(2,1) at 259 GHz is shown in Fig. \ref{fig.HDCO_mom1}. The compact disk emission is clearly seen. However, in contrast to the H$_2$CO emission in Fig. \ref{fig.H2CO_cont}, the full extended structure is not visible in this map. Fig. \ref{fig.HDCO_mom1}, which shows a moment 1 map integrated between the same velocity range as Fig. \ref{fig.streamcont}, confirms that of the two velocity structures seen in H$_2$CO (see Fig. \ref{fig.streamcont}), only the more compact velocity structure representing disk emission is fully present in HDCO. However, there is some asymmetry present in the morphology of this transition that would not be expected from the disk. Through comparison with Fig. \ref{fig.streamcont}, it is possible that this is arising at the base of the dynamic feature in the redshifted regime. 
\begin{figure}[!h]
\centering
\includegraphics[width=230pt]{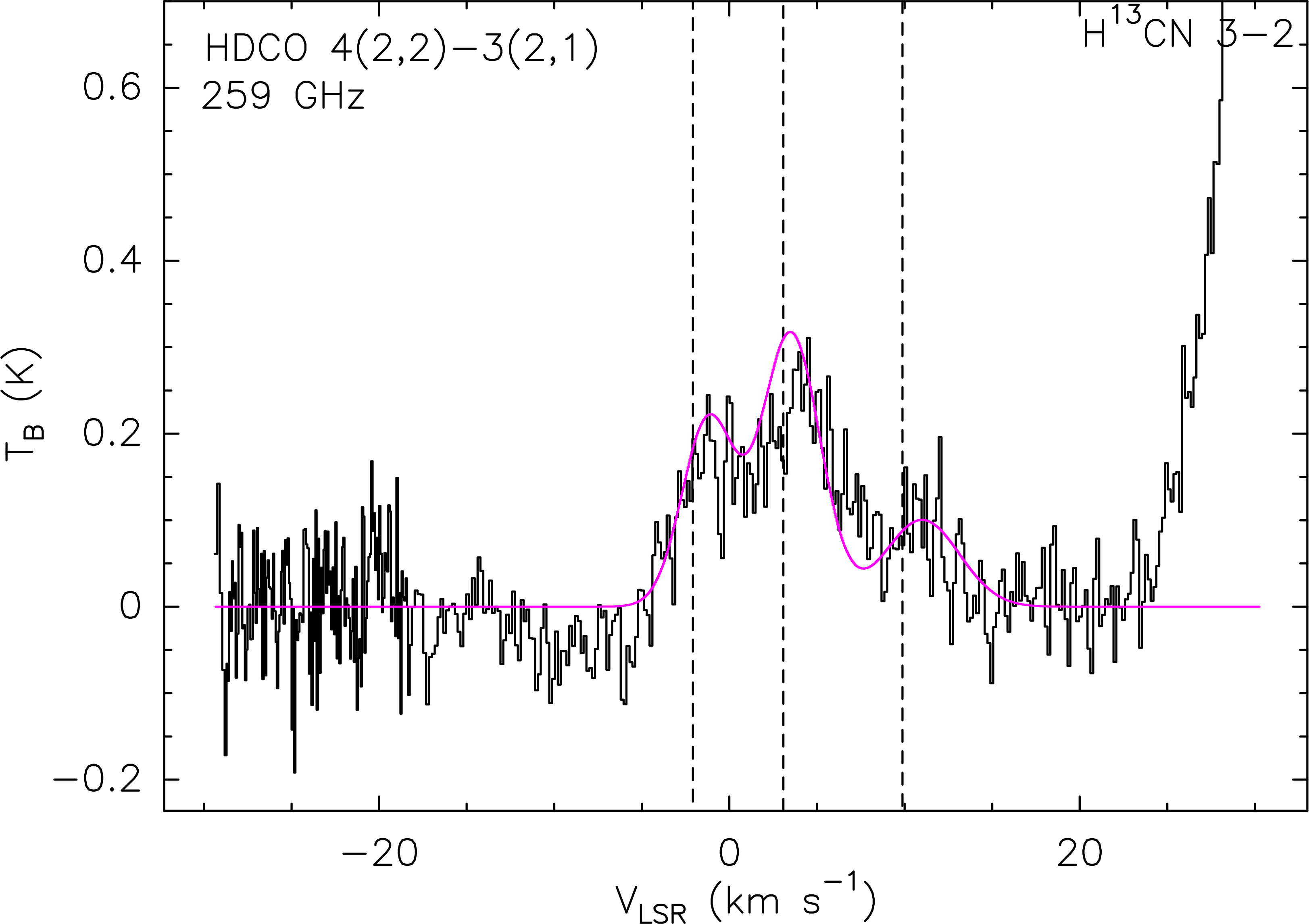}
\caption{HDCO spectrum (black) with velocity resolution 0.14 km~s$^{-1}$ for the 4(2,2)-3(2,1) transition overlaid with the best fit (pink) from our simultaneous radiative transfer analysis, extracted within the 4$\sigma$ contour shown in Fig. \ref{fig.HDCO_mom1}. The analysis was run twice, assuming upper limits for the size of the emitting size of 1.0" and 0.3" for Sources A and B (respectively) and lower limits of 0.5" and 0.15" for Sources A and B (respectively); the figure shown represents the latter. Dashed lines denote the previously identified velocity components at -2, 2.8 and 9.9 km~s$^{-1}$.}
    \label{fig.HDCOgauss}
\end{figure}
\\
\\
As can be seen in Fig. \ref{fig.HDCOgauss}, we extracted the HDCO 4(2,2)-3(2,1) transition, although this spectral window (spw) also contains a much stronger transition of H$^{13}$CN (3-2). However, this transition (rest frequency 259.0118 GHz) has a central peak that is well separated from the HDCO transition (rest frequency 259.0349 GHz) and therefore does not interfere at all with our analysis. The HDCO transition shows three emission components between the velocities of $\sim$ -5 to 12 km~s$^{-1}$, which likely originate from the disk. Indeed, as shown by the contours of Fig. \ref{fig.HDCO_mom1}, parts of the HDCO emission, notably that contributing to the aforementioned asymmetric morphology, fall outside the 4$\sigma$ detection threshold denoted by the cyan contour and so this supports the notion that only disk emission is included in the extracted spectrum seen in Fig. \ref{fig.HDCOgauss}. More specifically, comparing Fig. \ref{H2CO_3mom0}, which shows the morphology of each velocity component relative to Sources A and B, to Fig. \ref{fig.HDCOgauss}, the HDCO component between $\sim$ -5 to 0 km~s$^{-1}$ and the HDCO component with a lower limit of 6 km~s$^{-1}$ are likely originating from Source B while the HDCO component between $\sim$ 1 to 6 km~s$^{-1}$ are likely from Source A, similarly to the H$_2$CO emission. However, the elevated baseline visible in this spw makes defining the upper limit of the velocity range of the emission difficult. The spectrum is flat between $\sim$ 12 to 20 km~s$^{-1}$ but is consistently at an intensity of approximately 0.1 K. The flatness of the spectrum seemed to be too consistent for any lines to be present, therefore, we postulated that this may be an effect of the wing emission of the nearby strong H$^{13}$CN transition. In order to check this, the spw was re-cleaned using the same method (the FAUST pipeline) but with a reduced velocity window excluding the H$^{13}$CN transition. In this way we can see whether this elevated baseline is still present when the H$^{13}$CN transition is disregarded, by only considering velocities $\sim$15 km~s$^{-1}$ either side of the centre. When comparing the narrower spectrum to the spectrum covering the full $\sim$50 km~s$^{-1}$ velocity range (shown in Fig. \ref{fig.HDCOgauss}), the baseline around 12 to 20 km~s$^{-1}$ is still at a consistent intensity of 0.1 K. Additionally, not only does the frequency range of continuum subtraction cover this spw but also there was no evidence of issues of bandpass calibration which could have caused such an artefact to appear. Thus, we conclude that the cause of the elevated baseline is most likely line emission of HDCO. However, Fig. \ref{fig.HDCO_mom1} is integrated between a velocity range that excludes the affected component so that it is certain that only confirmed HDCO 4(2,2)-3(2,1) emission is included, while the spectrum shown in Fig. \ref{fig.HDCOgauss} has been corrected for this raised baseline.
\subsection{HDCO 4(1,4)-3(1,3)}
\begin{figure}[!h]
\centering
\includegraphics[width=230pt]{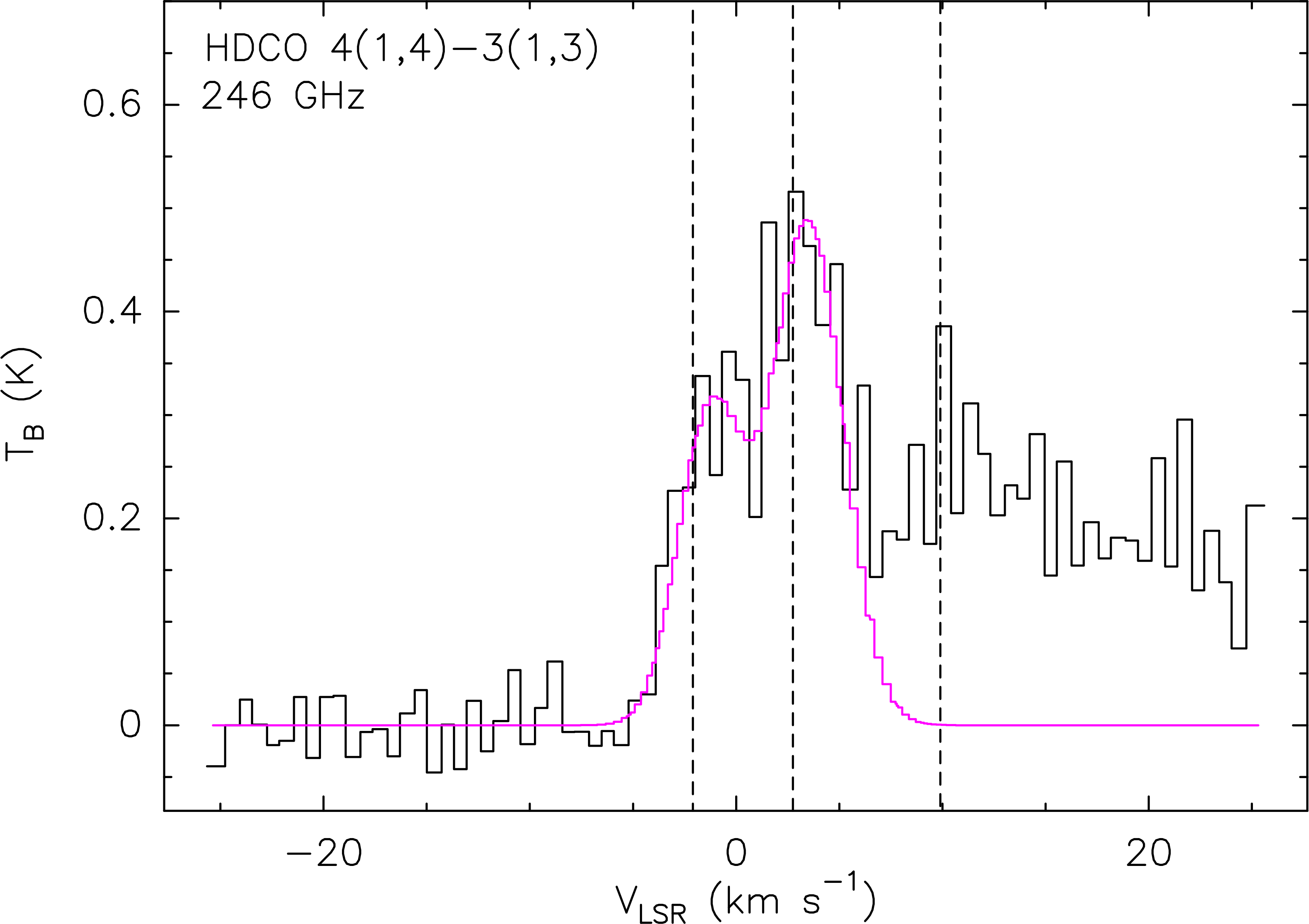}
\caption{HDCO spectrum (black) with velocity resolution 1.18 km~s$^{-1}$ for the 4(1,4)-3(1,3) transition overlaid with the best fit (pink) from our simultaneous radiative transfer analysis, extracted within the 4$\sigma$ contour shown in Fig. \ref{fig.HDCO_mom1}. The analysis was run twice, assuming upper limits for the size of the emitting size of 1.0" and 0.3" for Sources A and B (respectively) and lower limits of 0.5" and 0.15" for Sources A and B (respectively); the figure shown represents the latter. Note that the third component is not included in the fit due to blending with a transition of CH$_3$OCHO 19(4,15)-18(4,14) A, the emission components of which are clearly seen above 6 km~s$^{-1}$ and highlighted explicitly in Fig. \ref{fig.ch3ocho}. Dashed lines denote the previously identified velocity components at -2, 2.8 and 9.9 km~s$^{-1}$.}
    \label{fig.gauss}
\end{figure} 
\begin{figure}[!h]
    \centering
    \includegraphics[width=230pt]{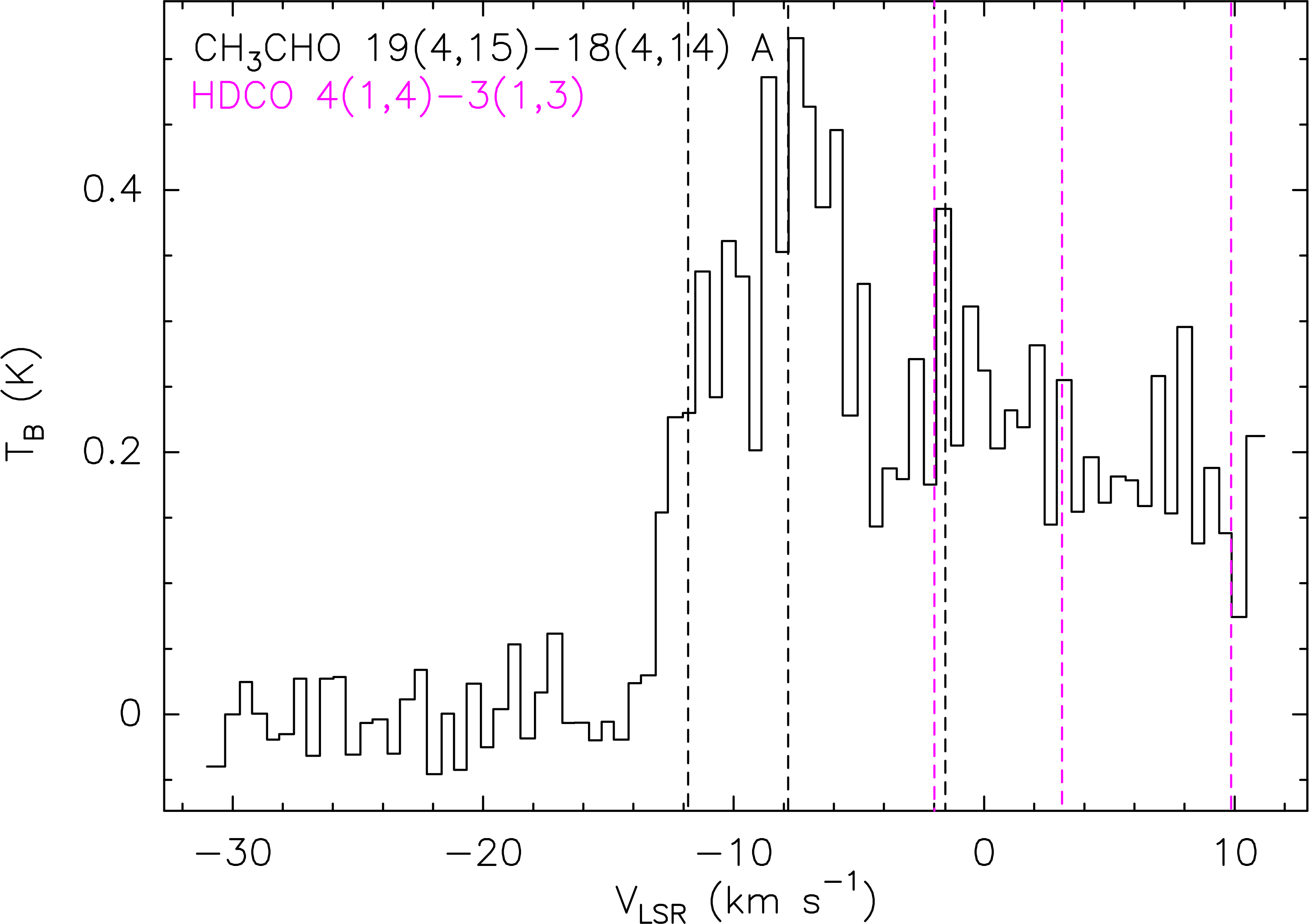}
    \caption{Extracted spectrum centred on the rest frequency of the CH$_3$OCHO 19(4,15)-18(4,14) A transition at 246.9147 GHz with dashed lines showing the predicted locations of the velocity components assuming similar V$_{\rm LSR}$ to methanol components (-2, 2.8 and 9.9 km~s$^{-1}$). The locations of the same velocity components (-2, 2.8 and 9.9 km~s$^{-1}$) in HDCO 4(1,4)-3(1,3) are shown by the pink dashed lines. The -2 km~s$^{-1}$ component of CH$_3$OCHO 19(4,15)-18(4,14) A is blended with the 9.9 km~s$^{-1}$ component of HDCO 4(1,4)-3(1,3).}
    \label{fig.ch3ocho}
\end{figure}
\begin{figure}[!h]
    \centering
    \includegraphics[width=230pt]{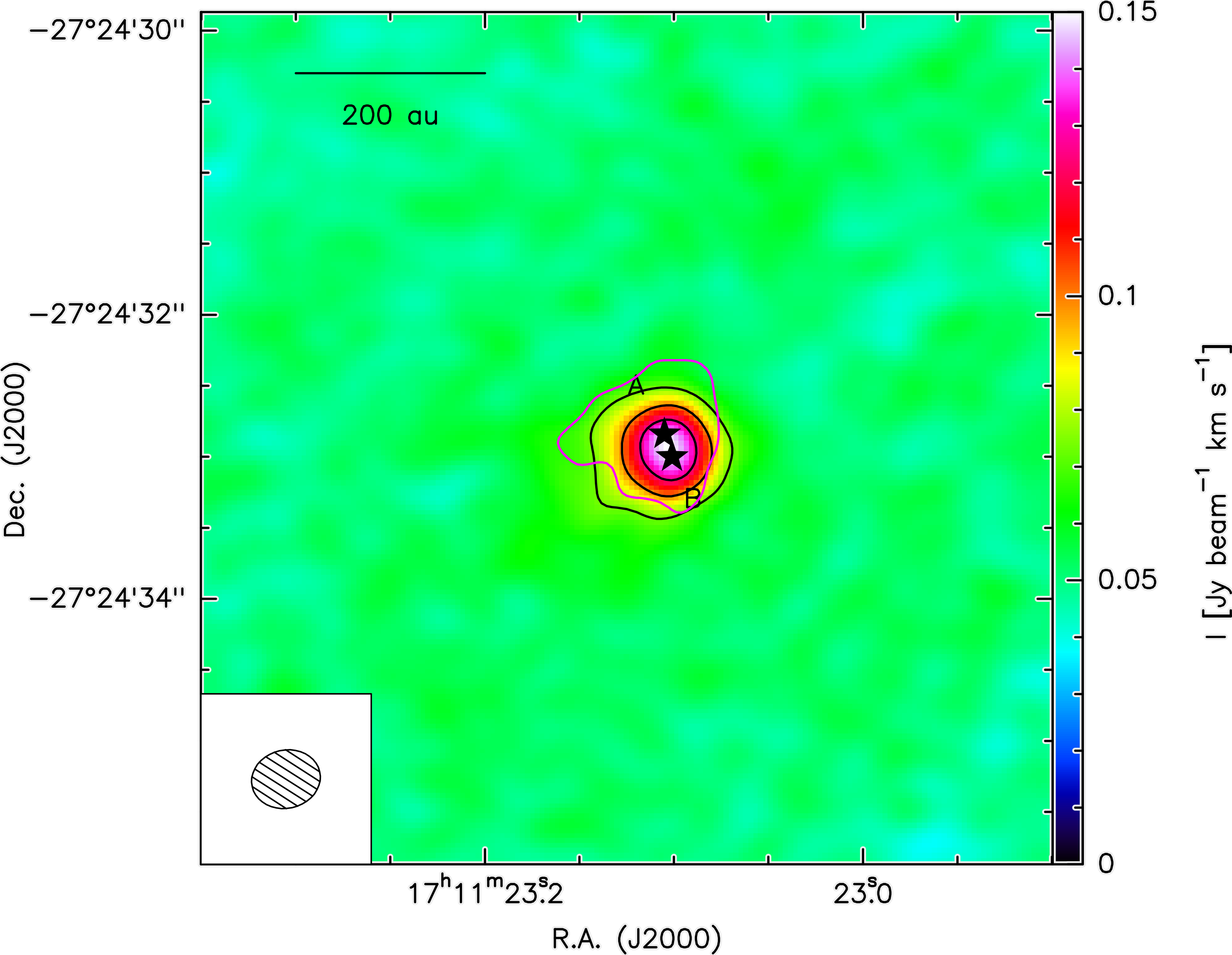}
    \caption{Moment 0 map of HDCO 4(1,4)-3(1,3) integrated between -5 and 8 km~s$^{-1}$. This velocity range was chosen to include the two unblended HDCO emission components shown in Fig. \ref{fig.gauss}. Contours start at 4$\sigma$ and end at 16$\sigma$ in 4$\sigma$ increments ($\sigma$=8.5 mJy~beam$^{-1}$~km~s$^{-1}$). The pink contour illustrates the area within which the spectrum in Fig. \ref{fig.gauss} was extracted, defined as the 4$\sigma$ contour shown in Fig. \ref{fig.HDCO_mom1}. The locations of protostars A and B are shown by the black stars. The ALMA synthesised beam is shown in the lower left corner.}
    \label{fig.HDCO_cont_mom0}
\end{figure}

\noindent As illustrated in Fig. \ref{fig.gauss}, the HDCO 4(1,4)-3(1,3) spectrum is also consistent with three emission components at $\sim$ -5 to 0 km~s$^{-1}$, $\sim$ 1 to 6 km~s$^{-1}$ and a third component with v$\rm_{LSR}~\geq$ 6 km~s$^{-1}$ that is affected by blending. The line identification software in CASSIS was used to check the transitions that could potentially cause this blending; based on the Einstein coefficients and $E_{\rm u}$ (upper energy level - using a maximum value of 150 K), the most likely culprit is CH$_3$OCHO 19(4,15)-18(4,14) A, as illustrated in Fig. \ref{fig.ch3ocho}. With the HDCO components showing similar velocity values to those seen in H$_2$CO (see Fig. \ref{H2CO_3mom0}), CO (\citealt{Alves17}) and CH$_3$OH (\citealt{vastel}), we conclude that, similar to these species, the high velocity HDCO emission components are most likely originating from Source B while the central component likely arises from Source A.
\\
\\
Fig. \ref{fig.HDCO_cont_mom0} shows the moment 0 map of the HDCO 4(1,4)-3(1,3) emission, along with the area used to extract the spectra seen in Figs. \ref{H2CO_3mom0}, \ref{fig.HDCOgauss} and \ref{fig.gauss}. The emission is compact and, unlike the HDCO 4(2,2)-3(2,1) emission, symmetrical, suggesting that this emission is likely originating only from the disk.
\subsection{D$_2$CO 4(0,4)-3(0,3)}
\begin{figure}
\centering
\includegraphics[width=230pt]{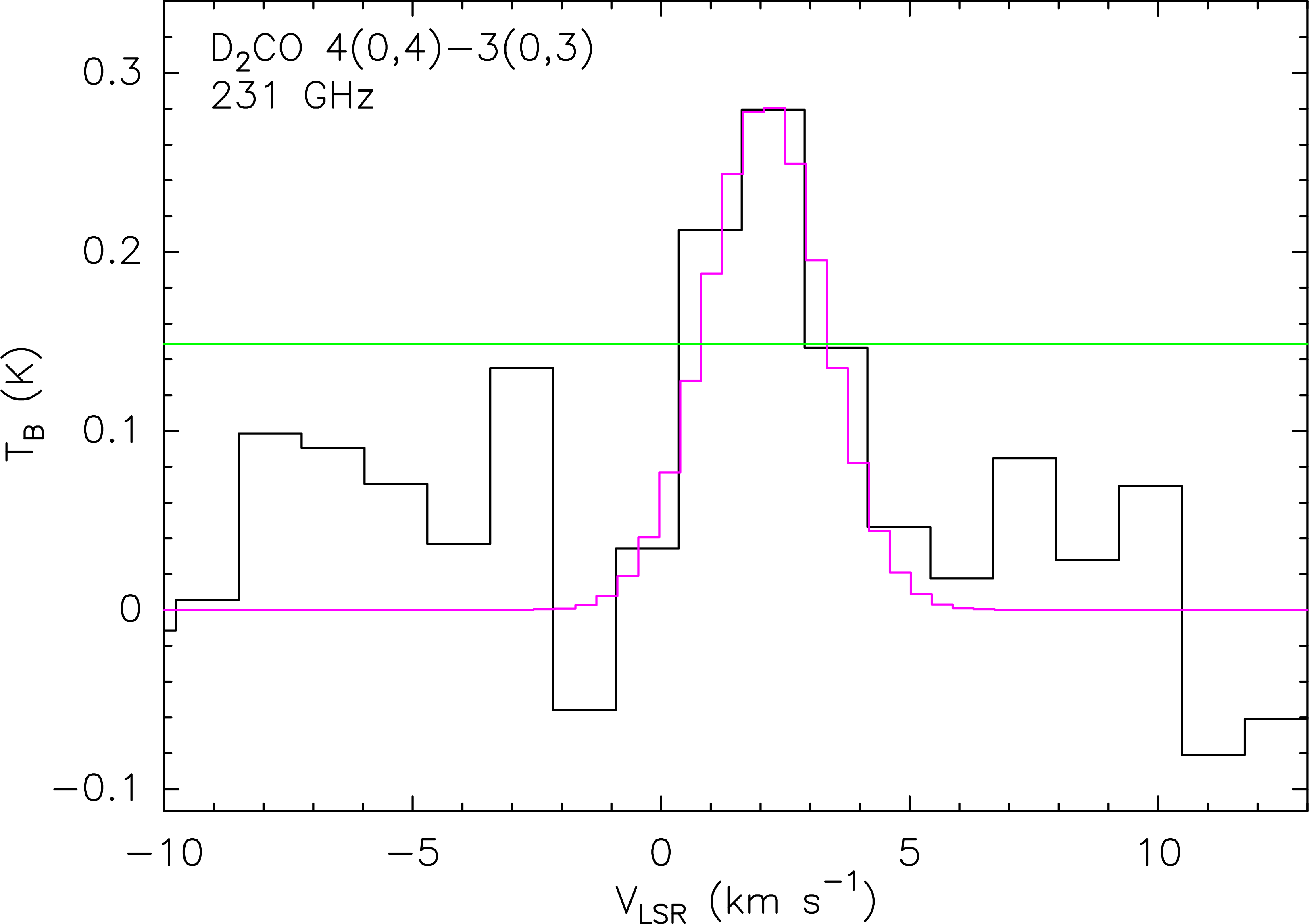}
\caption{D$_2$CO 4(0,4)-3(0,3) spectrum (black) with velocity resolution 0.16 km~s$^{-1}$ extracted from the 4$\sigma$ contour shown in Fig. \ref{fig.HDCO_mom1} and smoothed three times with the Hanning algorithm overlaid with best fit (pink) obtained using single component Gaussian fitting with FWHM, V$_{\rm{LSR}}$ and $T_{\rm{ex}}$ constrained using values obtained from fitting the HDCO transitions. The analysis was run twice, assuming upper limit values for the size of the emitting size of 0.3" (covering source B emission) and 1.0" (covering source A emission); the figure shown represents the latter. The \textit{rms} level (0.15 K) is shown in green.}
    \label{fig.D2CO_fit}
\end{figure}
As can be seen in Fig. \ref{fig.D2CO_fit}, a single peak is seen in D$_2$CO 4(0,4)-4(0,3), with emission seen between approximately 1 and 4 km~s$^{-1}$. Since this peak is visibly weak, we use Eq. \ref{eq.intins} to calculate the integrated intensity (\textit{W}; K~km~s$^{-1}$) value for the 3$\sigma$ upper limit and compare this to the strength of our observations.
\begin{gather}
W=(3\times rms\times FWHM)\sqrt{(2\times cal)^2+\Bigg(2\times\frac{\Delta\rm{v}}{FWHM}\Bigg)}
\label{eq.intins}
\end{gather}
\noindent In this equation, \textit{cal} represents the calibration error (taken in this case to be 10\%), $\Delta{\rm v}$ is the velocity resolution (which is 0.2 km~s$^{-1}$) and \textit{rms} is 0.15 K. Therefore, using a measured FWHM linewidth obtained using Gaussian fitting within CASSIS as 1.4 km~s$^{-1}$, we obtain a 3$\sigma$ upper limit value of 0.36 K~km~s$^{-1}$. The integrated intensity of our D$_2$CO line is 0.826 K~km~s$^{-1}$, therefore yielding a ratio of 2.3 compared to the 3$\sigma$ threshold. Thus, our D$_2$CO observations have a s/n ratio of approximately 7$\sigma$ in integrated intensity.
\subsection{Emission Origin}\label{origin}
As introduced in Sect. \ref{Introduction} (see also \citealt{rodgers}), the grain surface formation pathways for methanol and formaldehyde are intrinsically linked in that both form through hydrogenation reactions on the surface itself:
\begin{gather}
\rm{CO + H} \rightarrow \rm{HCO},\\
\rm{HCO + H} \rightarrow \rm{H_2CO},\\
\rm{H_2CO + H} \rightarrow \rm{CH_3O},\\
\rm{CH_3O + H} \rightarrow \rm{CH_3OH}.
\end{gather}
During any of these stages, it is possible for a D-addition reaction to occur rather than H-addition, creating a grain surface deuteration pathway. It is widely understood (\citealt{watanabe02}, \citealt{rimola}) that these grain surface formation processes occur at low temperatures during the cold prestellar phase during which the atoms and molecules are frozen in the icy mantles of dust grains. This is followed by subsequent injection of the particles into the gas phase due to the increase in temperature associated with the protostellar phase. More specifically, this temperature rise can be the result of thermal (e.g. the presence of hot corinos - \citealt{taquet16}) or non-thermal (e.g. shocks caused by interaction of material with quiescent gas - \citealt{jiminez}) processes. Previous analysis of methanol (\citealt{vastel}) and formaldehyde (\citealt{alves_19}) emission towards [BHB2007] 11 has shown that the dust is optically thick towards the protobinary system and that the emission of both formaldehyde and methanol shows an increasing velocity gradient towards protostar B. This evidence strongly favours the scenario in which the injection of formaldehyde and methanol into the gas phase has occurred due to a shock as the observations appear to rule out the suggestion that A and/or B are hot corinos; indeed, \citet{vastel} found that the CH$_3$OH emission peak appears spatially shifted with respect to both sources. In addition, dynamic material, which may be involved in causing such a shock, has been observed in this source previously (\citealt{alves_19}), as well as in the present work.
\begin{figure}[H]
    \centering
    \includegraphics[width=230pt]{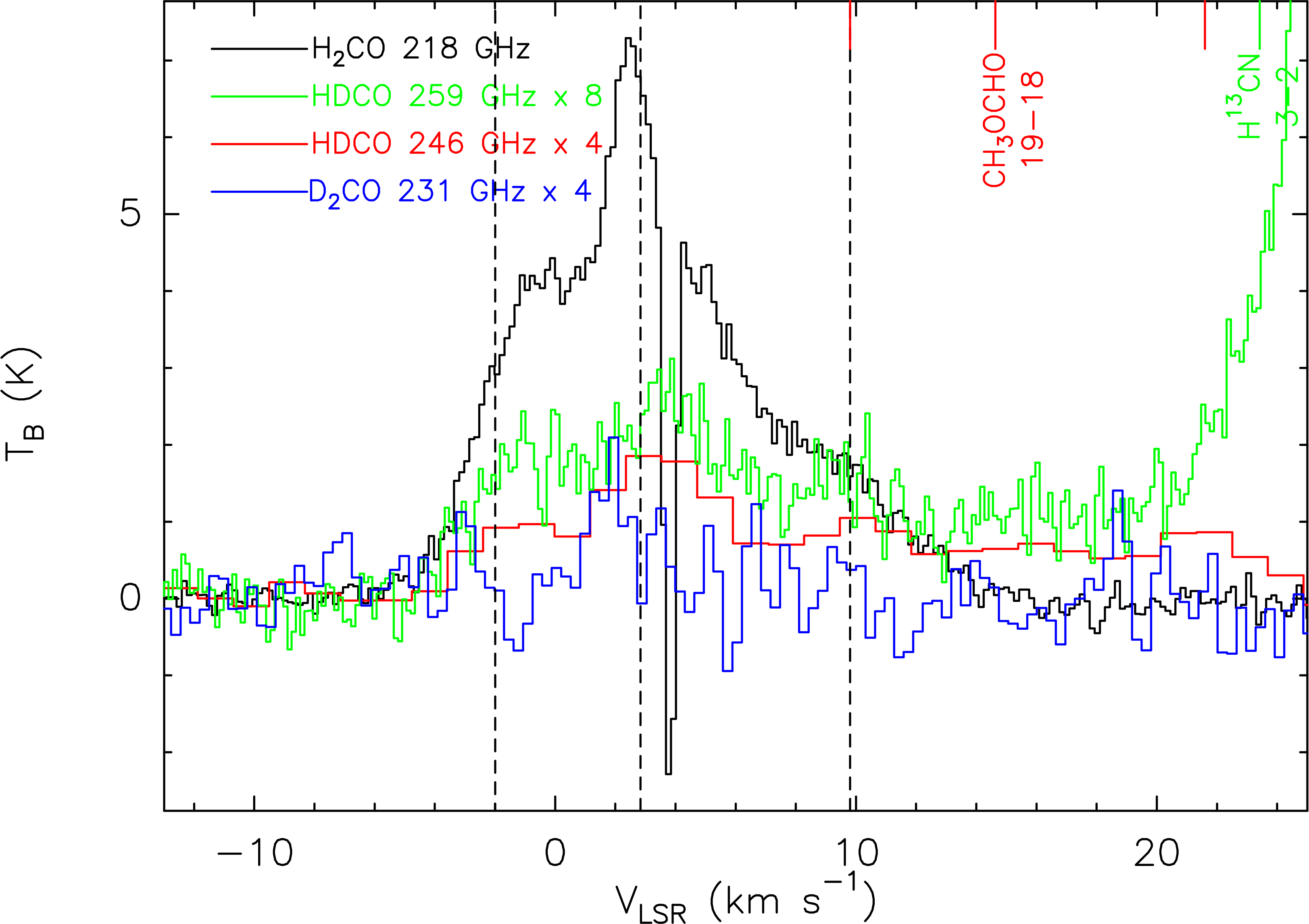}
    \caption{Comparison of H$_2$CO 3(0,3)-2(0,2), D$_2$CO 4(0,4)-3(0,3), HDCO 4(1,4)-3(1,3) and HDCO 4(2,2)-3(2,1) lines with the baselines subtracted
    . The dashed lines represent velocities of -2, 2.8 and 9.9 km~s$^{-1}$. The CH$_3$OCHO 19(4,15)-18(4,14) A transition (246.9147 GHz) that is near HDCO 4(1,4)-3(1,3) is shown in red, while the H$^{13}$CN (3-2) transition (259.0118 GHz) in the same spw as HDCO 4(2,2)-3(2,1) is denoted in green.}
    \label{fig.H2CO_overlap}
\end{figure}
\noindent Fig. \ref{fig.H2CO_overlap}, which overlaps our four spectra on a single plot, also shows dashed lines denoting the three velocity components first identified in CH$_3$OH emission towards [BHB2007] 11 as described in \citet{vastel} at -2, 2.8 and 9.9 km~s$^{-1}$. Notably, this previous research suggests that the -2 and 9.9 km~s$^{-1}$ components are originating from higher velocity warm gas streaming towards Source B (see also \citealt{alves_19}, who resolved these components for the first time), 
while the 2.8 km~s$^{-1}$ CH$_3$OH component is believed to originate from Source A. Fig. \ref{fig.H2CO_overlap} also shows multiple velocity components in H$_2$CO and HDCO, which show consistency with the three components first identified in CH$_3$OH emission by \citet{vastel}. Meanwhile, D$_2$CO shows a single peak with a velocity shift compared to the previously identified velocity components. 
\\
\\
Comparing our H$_2$CO observations with the high resolution CH$_3$OH observations presented in \citet{vastel}, clear differences are apparent in the morphology of the emission of these species. Notably, it is very clear that CH$_3$OH is very compact compared to the more extended H$_2$CO seen in the present observations. This is despite the fact that both species are injected into the gas phase from the frozen surfaces of dust grains due to the rising temperature of the system under similar conditions. The $T_{\rm ex}$ values obtained from non-LTE analysis of the CH$_3$OH emission ranged from 110-130 K, suggesting that this is tracing hot gas in this object (\citealt{vastel}). Infall motions suggest the presence of a shocked region which likely enhances the production of CH$_3$OH on small scales, while the low population of the upper states of the observed CH$_3$OH lines makes this species harder to detect in the cold envelope region. 
All of these factors may explain why CH$_3$OH and H$_2$CO appear to be tracing different regions of [BHB2007] 11, although it is also possible to explain the observed morphology differences if H$_2$CO is also being produced in the gas phase.
\\
\\
Fig. \ref{fig.pixels_map} shows the moment 1 map for H$_2$CO emission in colour scale overlaid with five pixel positions from which we extract individual spectra. As previously discussed, there appears to be separate velocity components along two different axes in this moment 1 map, with the more east-west aligned components showing lower velocities than the components aligned with the continuum emission extension. Four of the pixels shown represent the blue- and redshifted velocity components along both axes, while the fifth represents the central position and the peak of the emission. Each pixel outside the centre was chosen as the most intense within four velocity ranges: 0-2.5 km~s$^{-1}$, 2.5-3.5 km~s$^{-1}$, 3.5-4 km~s$^{-1}$ and 4-6 km~s$^{-1}$. The H$_2$CO, HDCO (both transitions) and D$_2$CO spectra extracted from these five pixels are shown in Fig. \ref{fig.pixels}. It is clear from these spectra that, while the H$_2$CO emission is extended, the HDCO emission is compact and not seen in any of the spectra outside that extracted from the central pixel; indeed, HDCO is only seen within $\sim$ 300 au of the centre. From this, we postulate that HDCO likely originates from the shocked regions first identified in \citet{vastel} located to both sides of Source B and also near Source A. Moreover, the absorption feature is only seen in the H$_2$CO spectrum extracted from the central pixel, further illustrating that this feature is not seen outside $\sim$ 500 au. Meanwhile, the low s/n of the D$_2$CO spectra in these particular pixels prevents any conclusions from being drawn.
\begin{figure}[!h]
    \centering
    \includegraphics[width=230pt]{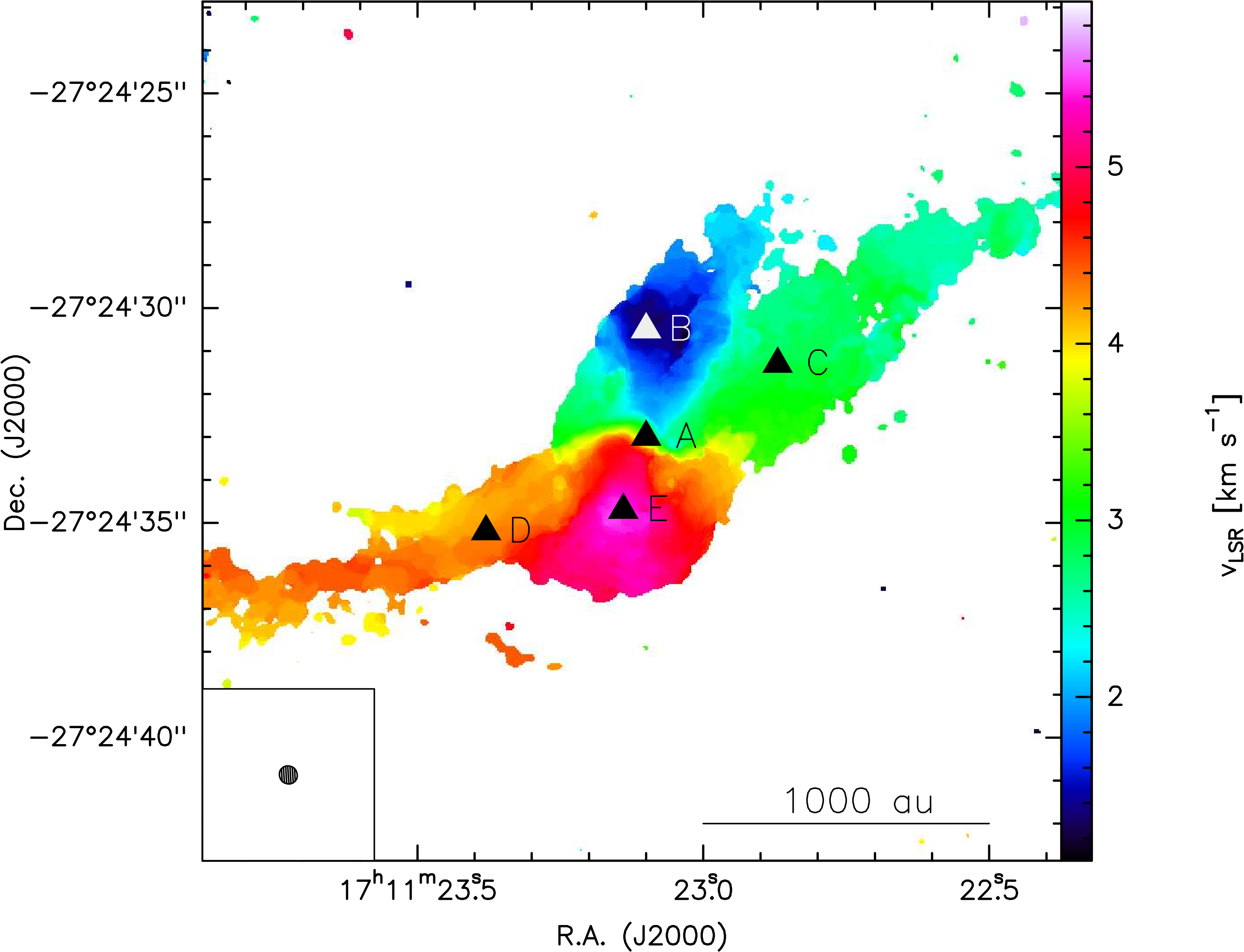}
    \caption{H$_2$CO moment 1 map in colour scale with five pixels overlaid in black and white (for clarity in this case) triangles, chosen as the pixel at the location of the dust peak (A), along with the most intense pixels for each velocity range (B: 0-2 km~s$^{-1}$, C: 2.5-3.5 km~s$^{-1}$, D: 3.5-4 km~s$^{-1}$ and E: 5-6 km~s$^{-1}$). The ellipse in the bottom-left corner represents the ALMA synthesised beam.}
    \label{fig.pixels_map}
\end{figure}
\begin{figure*}[!h]
    \centering
    \captionsetup[subfigure]{labelformat=empty}
    \begin{subfigure}{}
    \includegraphics[width=215pt]{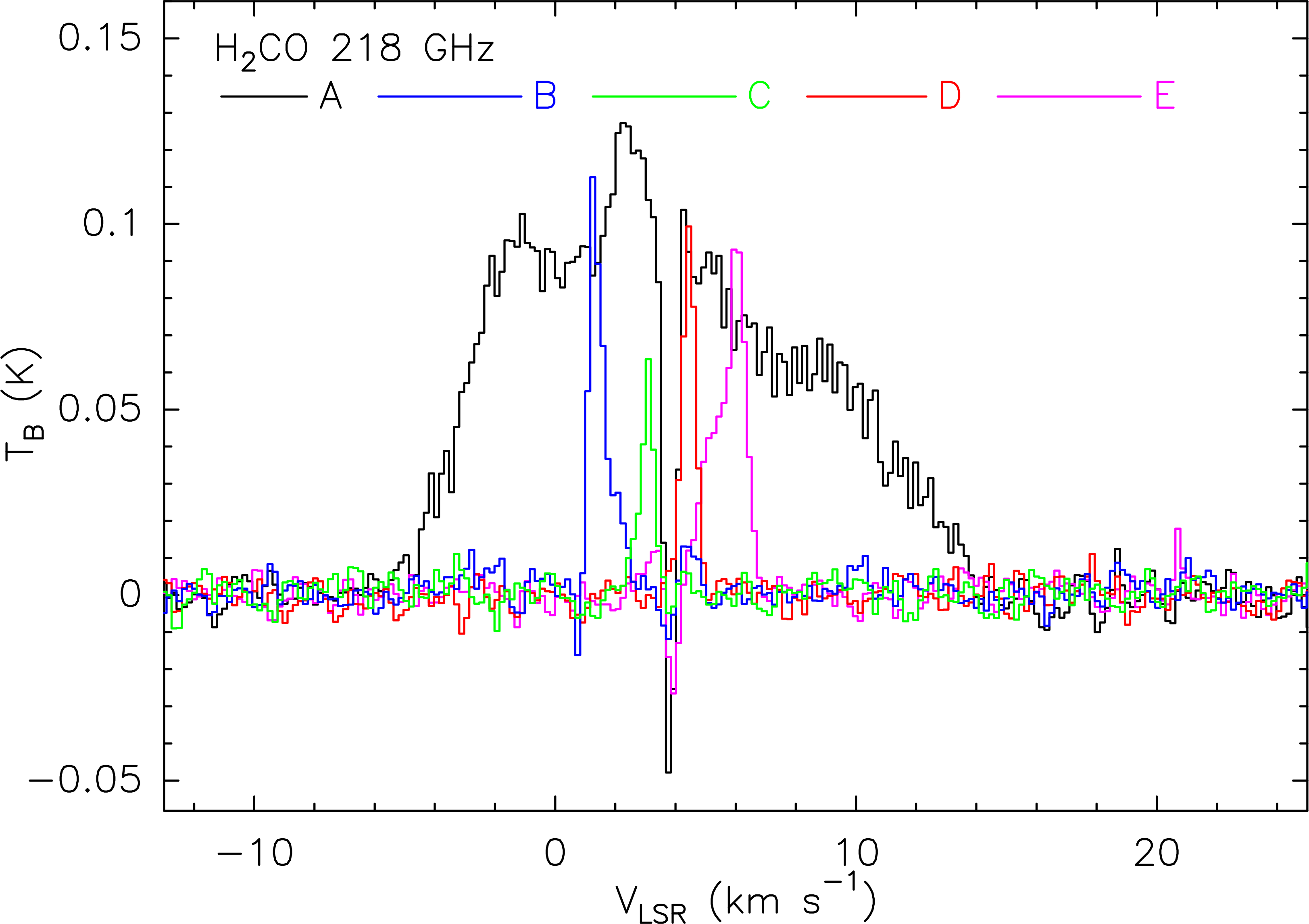}
    \end{subfigure}
    \begin{subfigure}{}
    \includegraphics[width=230pt]{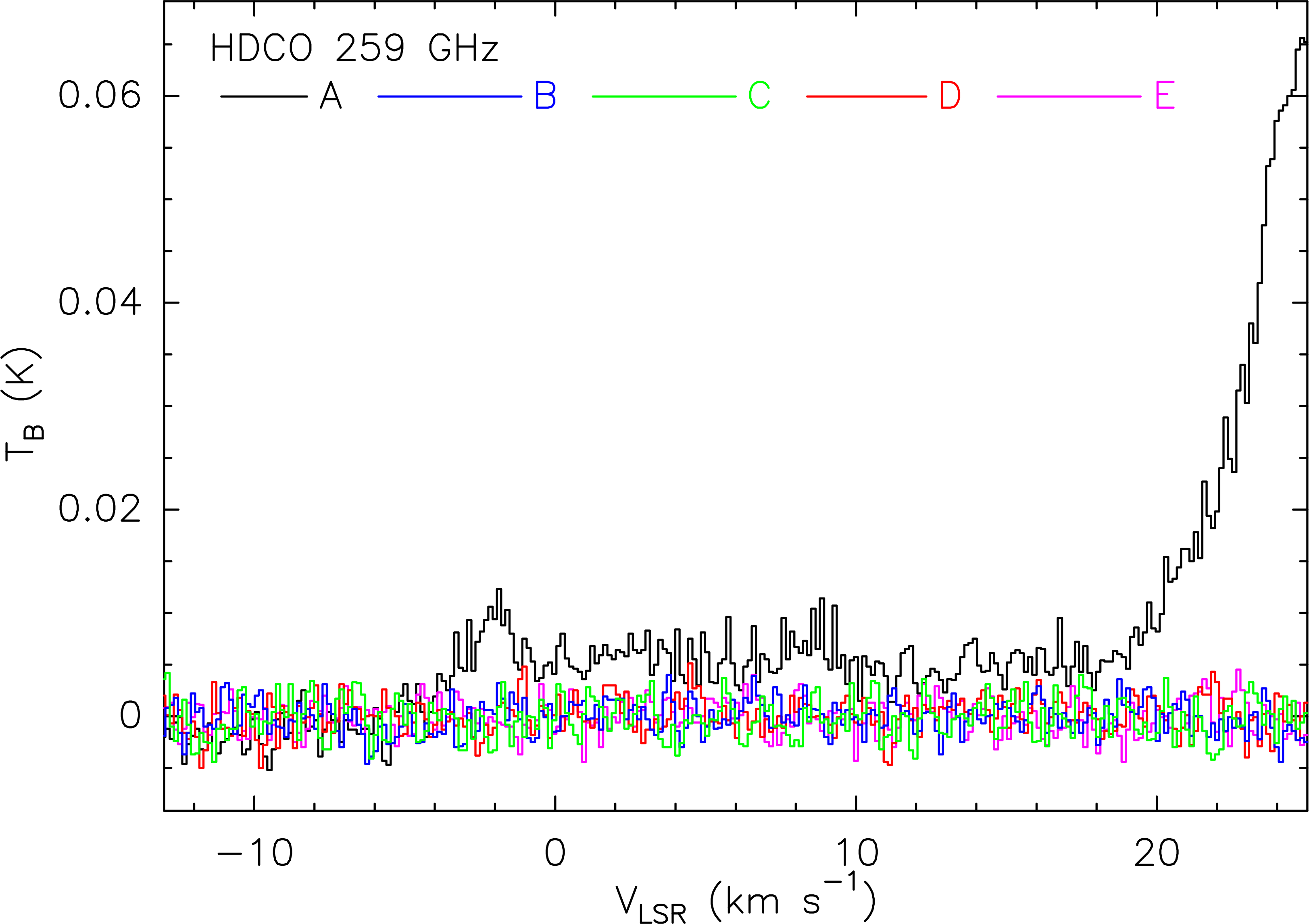}
    \end{subfigure}
    \begin{subfigure}{}
    \includegraphics[width=230pt]{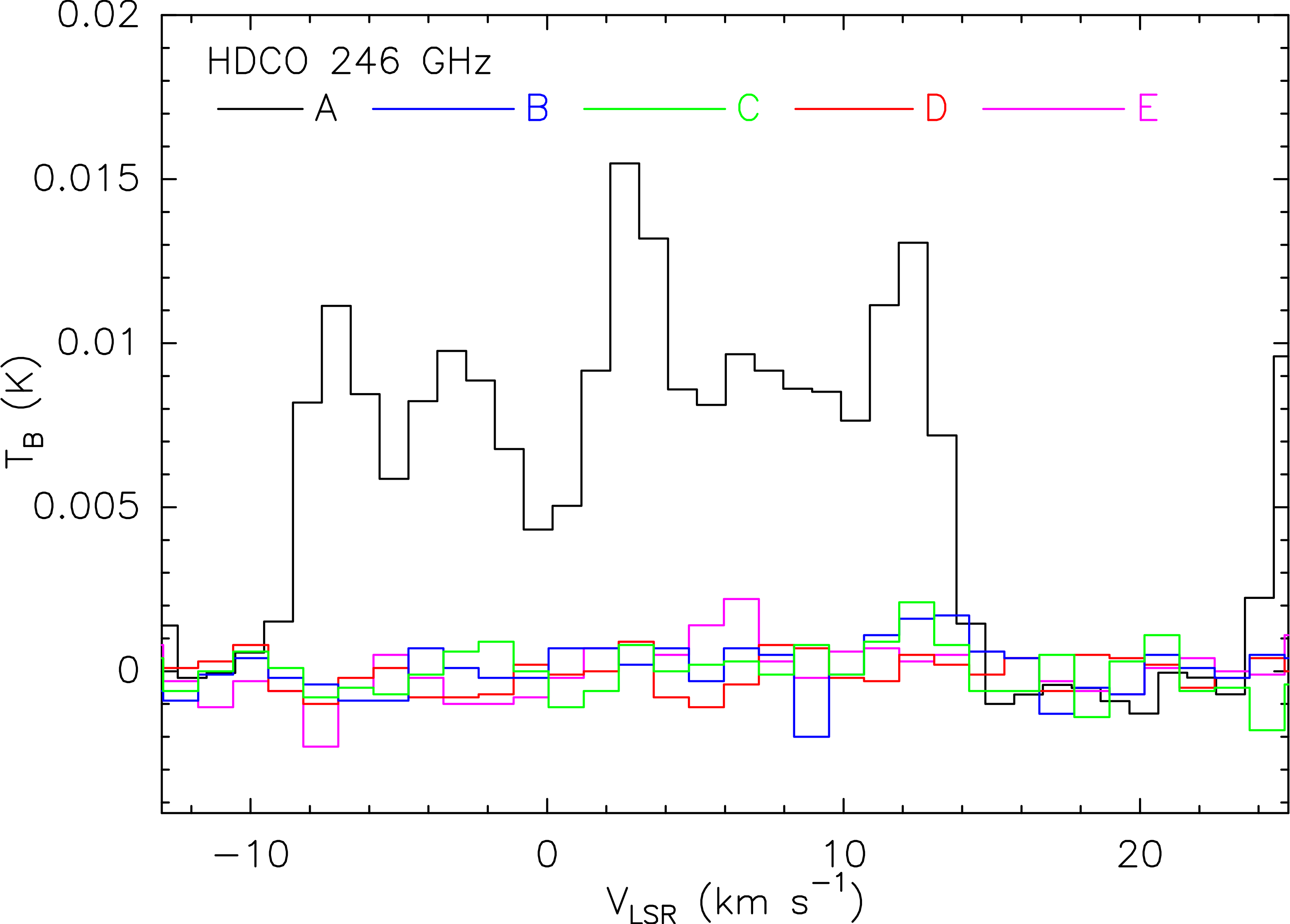}
    \end{subfigure}
    \begin{subfigure}{}
    \includegraphics[width=230pt]{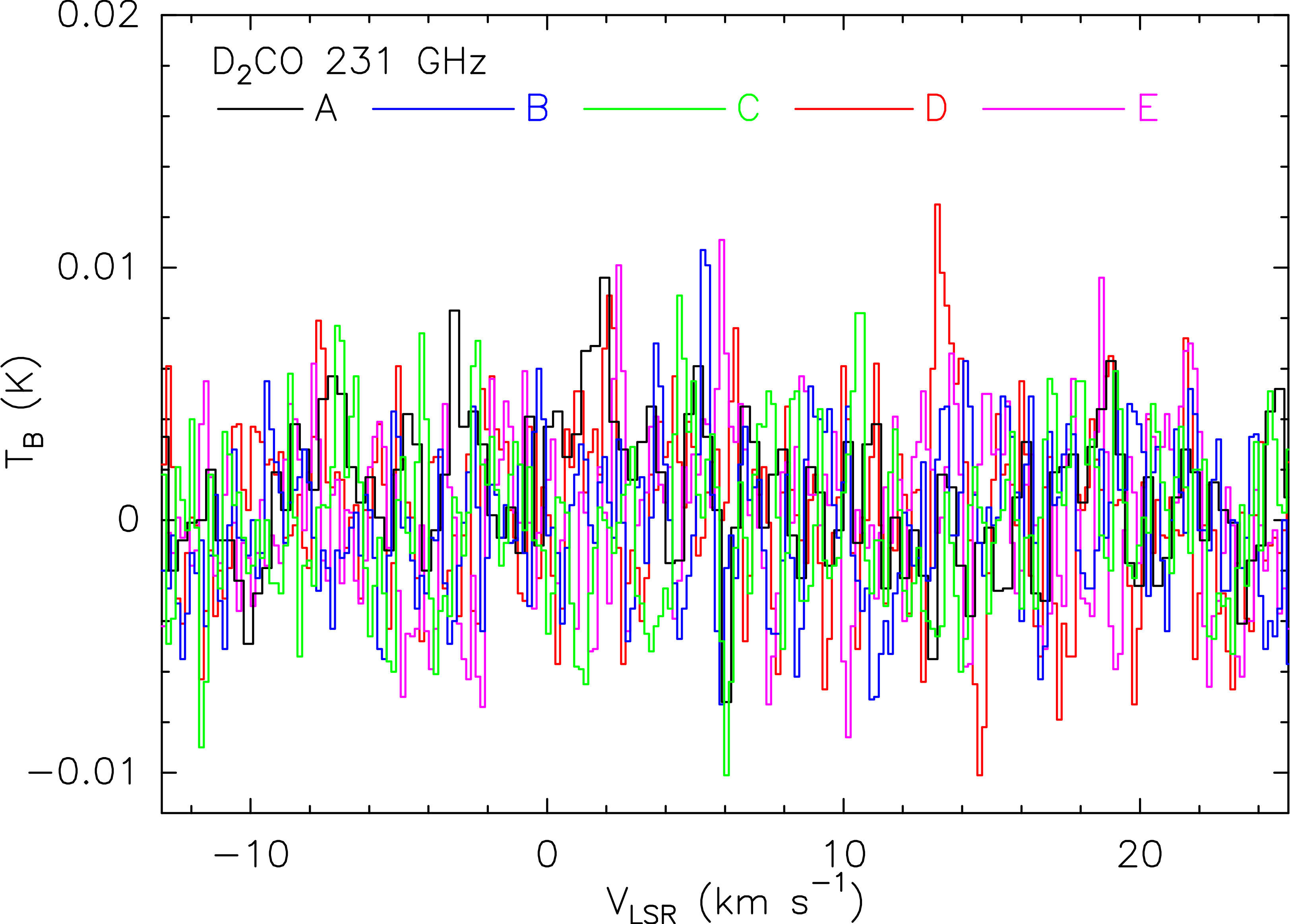}
    \end{subfigure}
    \caption{Spectra extracted from the five pixels (labelled A, B, C, D and E) indicated in Fig. \ref{fig.pixels_map}. Clockwise from upper left: H$_2$CO (218 GHz), HDCO (259 GHz), D$_2$CO (231 GHz) and HDCO (246 GHz).}
    \label{fig.pixels}
\end{figure*}
\section{Radiative Transfer Analysis}\label{radtrans}
The molecular column densities have been derived assuming Local Thermodynamic Equilibrium (LTE) conditions. Therefore, we can utilise equations from the Formalism for CASSIS Software, specifically within the "Documentation and Data" section. In particular, we use Eq. \ref{eq.ncol} for the calculation of the total column density ($N_{\rm TOT}$):
\begin{equation}
N_{\rm TOT}=\frac{8\pi\nu^3(\Delta \rm{V})\tau_0 Q(T_{\rm ex})\sqrt{\pi}}{2\sqrt{ln2}}\frac{e^{\frac{E_{\rm u}}{T_{\rm ex}}}}{c^3A_{\rm ul}g_{\rm u}(e^{\frac{h\nu}{kT_{\rm ex}}}-1)},
\label{eq.ncol}
\end{equation}
where \textit{c}, \textit{h} and \textit{k} represent the speed of light, Boltzmann constant and Planck constant, respectively, in cgs units, while $\tau_0$ is the opacity at the centre of the line. We take the background temperature (\textit{T$_{\rm bg}$}) as that of the cosmic microwave background (CMB; i.e. 2.73 K). We obtain the FWHM ($\Delta$V in cm~s$^{-1}$ in Eq. \ref{eq.ncol}) through Gaussian fitting of the spectra. Frequency ($\nu$ in Hz),  partition function (Q($T_{\rm ex}$)), Einstein coefficient ($A_{\rm ul}$), energy of the upper level ($E_{\rm u}$ in K) and upper level population ($g_{\rm u}$) are taken from the Cologne Database for Molecular Spectroscopy (CDMS; \citealt{muller01}).
\subsection{HDCO}\label{HDCO}
Since there are two HDCO transitions in this sample, simultaneous Levenberg-Marquardt fitting within CASSIS of both transitions was employed in order to obtain the initial Gaussian fit of the extracted spectra, yielding best fit values for V$_{\rm LSR}$, FWHM and intensity for each visible component. However, as previously mentioned, one of the components in HDCO 4(1,4)-3(1,3) shows blending with one of the components of CH$_3$OCHO 19(4,15)-18(4,14) A, as predicted in Fig. \ref{fig.ch3ocho}. Due to this, the two unblended components were fitted simultaneously with the corresponding components in HDCO 4(2,2)-3(2,1) in order to constrain FWHM and V$_{\rm LSR}$; the FWHM and V$_{\rm LSR}$ values for the blended component were thus fixed by fitting the 4(2,2)-3(2,1) transition alone; we apply the same FWHM and V$_{\rm LSR}$ values to the same component of both transitions in order to calculate $N_{\rm TOT}$ for this component. We also use the assumption that the HDCO emission is of a similar origin to that of CH$_3$OH to constrain our FWHM and V$_{\rm LSR}$ values. In order to check this assumption, we ran a 2D deconvolution of all three emission components in both HDCO transitions (with the exception of the blended component of the HDCO 4(1,4)-3(1,3) transition) applying the CASA \textit{imfit} command to the peak emission channel in each case. The results of this deconvolution returned image component size estimates (major axis FWHM x minor axis FWHM) of 0.34" $\times$ 0.27" for the component at approximately -1 km~s$^{-1}$, 1.37" $\times$ 0.68" for the component at approximately 3 km~s$^{-1}$ and 0.30" $\times$ 0.20" for the component at approximately 9 km~s$^{-1}$. These findings agree with the 2D deconvolution presented in \citet{vastel} in that the component associated with Source A is slightly more extended than those associated with Source B. Using these results, along with the consistency between emission components of CH$_3$OH and H$_2$CO, we therefore ran two LTE radiative transfer analyses, one assuming a lower limit emitting size of 0.15" (for the components associated with Source B) and 0.50" (for the Source A component) and the other assuming an upper limit emission area of 0.30" (for the components associated with Source B) and 1.00" (for the Source A component). 
Thus, all $N_{\rm TOT}$ values for HDCO are corrected for beam dilution accordingly.
\\
\\
Figs. \ref{fig.HDCOgauss} and \ref{fig.gauss} in Sect. \ref{Methods} show the Gaussian fitting for the HDCO transitions. Within CASSIS, we used constraints provided by our initial Gaussian fitting in order to obtain best fit values for FWHM, V$_{\rm LSR}$, $T_{\rm ex}$ and $N_{\rm TOT}$ of our two unblended components at $\sim$ -1 and $\sim$ 3 km~s$^{-1}$ using Monte Carlo Markov-Chain (MCMC) fitting, assuming a 10\% calibration error. As we assume a common origin of the disk emission, V$_{\rm LSR}$, FWHM and $T_{\rm ex}$ are also applicable to our H$_2$CO transition, enabling us to also obtain $N_{\rm TOT}$ values for this, which are reported in Table \ref{table_gauss}. Since we are assuming LTE, all excitation temperatures for one molecule are the same; therefore, these best fit parameters were then used to obtain an initial constraint for excitation temperature ($T_{\rm ex}$) for the HDCO component at $\sim$ -1 km~s$^{-1}$, which was found to be approximately 90 $^{+3}_{-8}$ K. 
\\
\\
Following our analysis, we also obtained an upper limit column density for absorption in HDCO of 6.6$\times$10$^{11}$cm$^{-2}$. We obtained this by noting the maximum $T_{\rm ex}$ for which absorption is observed after adding a continuum value of 10.3 K - obtained by measuring the peak intensity of the continuum observations of our dataset (see Fig. \ref{fig.H2CO_cont}) - 
and running an LTE analysis at this temperature.
\subsection{H$_2$CO}\label{H2CO_radtrans}
As we are assuming LTE conditions, we can also apply our best fit $T_{\rm ex}$ value obtained via our HDCO analysis to fit our single H$_2$CO transition. Therefore, using a fixed $T_{\rm ex}$ of 90 K and our best fit V$_{\rm LSR}$ and FWHM values from HDCO, we constrained Gaussian fits for each of the three emission components visible in the H$_2$CO spectrum. In addition to this, the absorption feature also required its own Gaussian fitting. To achieve this, we added a value for the continuum of 10.3 K from our observations and assumed an excitation temperature of 13 K to simulate the temperature of the cold envelope in which the absorption was occurring. This excitation temperature is the upper limit obtained for absorption in HDCO and, as such, the obtained column density for H$_2$CO in the absorption component is also an upper limit. We use this temperature in order to facilitate calculation of upper limits for the column density and D/H ratios in this component, which are shown in Table \ref{table_gauss}. We also note that the V$_{\rm LSR}$ value associated with the absorption component corresponds with the systemic velocity of the source (3.6 km~s$^{-1}$). The resultant fitting of H$_2$CO can be seen in Fig. \ref{H2CO_3mom0} in Sect. \ref{H2CO}.
\subsection{D$_2$CO}
Levenberg-Marquardt fitting yields a FWHM of 2.8$\pm$0.6 km~s$^{-1}$ and a V$_{\rm LSR}$ of 2.1$\pm$0.3 km~s$^{-1}$ for the single detected D$_2$CO component. This V$_{\rm LSR}$ value is not consistent with those obtained for H$_2$CO and HDCO, therefore it is unclear which of the three identified emission components this D$_2$CO emission corresponds to. Furthermore, due to the weakness of the emission, the 2D deconvolution failed to converge, leading us to assume that the D$_2$CO emission has a similar emitting size to that of HDCO. Therefore, we used LTE analysis with the aforementioned V$_{\rm LSR}$ and FWHM values, an assumed excitation temperature of 90 K and an upper limit emitting size of both 0.3" (covering Source B emission) and 1.0" (covering Source A emission) to obtain upper limit values of N$_{\rm TOT}$ for the single D$_2$CO component. Thus, once again, we take the effects of beam dilution into account. 
\subsection{D/H Ratio Derivation}\label{Results}
Table \ref{table_gauss} shows the best fit parameters, constrained using Levenberg-Marquardt fitting and MCMC modelling, for FWHM, V$_{\rm LSR}$ and intensity, along with the resultant values for $N_{\rm TOT}$, all with their relative uncertainties, for all transitions in our sample.
\begin{table*}[!h]
    \centering
    \caption{Best fit parameters for the Gaussian fits of each component in each spectrum, along with the resultant $N_{\rm TOT}$ and calculated $N_{\rm TOT}$ and D/H ratios.}
    \begin{tabular}{c c c c c c c c}
        \hline
        Species & Size & $T_{\rm ex}$ & V$_{\rm LSR}$ & FWHM & $N_{\rm TOT}$ & $N_{\rm TOT}$/$N_{\rm TOT}$(H$_2$CO) & D/H ratio\\
         & " & K & km~s$^{-1}$ & km~s$^{-1}$ & cm$^{-2}$ & \% & \\
         \hline
         \hline
         & & & & & & \\
         & 0.15 & 93$^{+2}_{-7}$ & -1.2$\pm$0.1 & 3.5$\pm$0.1 & (1.4$^{+0.1}_{-0.1}$)$\times$10$^{13}$ & 4$^{+5}_{-1}$ & (2$^{+2}_{-1}$)$\times$10$^{-2}$ \\
         & & & & & & \\
         & 0.50 & 95$^{+1}_{-3}$ & 3.4$\pm$0.1 & 3.9$\pm$0.1 & (2.5$^{+0.2}_{-0.3}$)$\times$10$^{13}$ & 3$^{+2}_{-1}$ & (1.5$^{+0.9}_{-0.5}$)$\times$10$^{-2}$ \\
         & & & & & & \\
         & 0.15 & 90 & 10$\pm$1 & 5.0$\pm$0.5 & (8.8$^{+0.9}_{-0.9}$)$\times$10$^{12}$ & 3$^{+7}_{-2}$ & (1$^{+4}_{-1}$)$\times$10$^{-2}$ \\
         & & & & & & \\
         & 0.15 & $<$~13 & 3.6 & 0.40 & $<$~1.8$\times$10$^{11}$~$^{(a)}$ & $<$~0.3 & $<$~1.7$\times$10$^{-3}$ \\
         HDCO$^{(b)}$ & & & & & & \\
         & 0.30 & 96$\pm$3 & -1.2$\pm$0.1 & 3.5$\pm$0.1 & (1.1$^{+0.1}_{-0.1}$)$\times$10$^{13}$ & 3$^{+4}_{-1}$ & (2$^{+2}_{-1}$)$\times$10$^{-2}$ \\
         & & & & & & \\
         & 1.00 & 98$^{+1}_{-2}$ & 3.4$\pm$0.1 & 3.9$\pm$0.1 & (2.9$^{+0.2}_{-0.3}$)$\times$10$^{13}$ & 3$^{+2}_{-1}$ & (1.8$^{+0.9}_{-0.5}$)$\times$10$^{-2}$ \\
         & & & & & & \\
         & 0.30 & 90 & 10$\pm$1 & 5.0$\pm$0.5 & (8$^{+1}_{-1}$)$\times$10$^{12}$ & 3$^{+7}_{-1}$ & (1$^{+4}_{-1}$)$\times$10$^{-2}$ \\
         & & & & & & \\
         & 0.30 & $<$~13 & 3.6 & 0.40 & $<$~6.6$\times$10$^{11~(a)}$ & $<$~1.3 & $<$~6.6$\times$10$^{-3}$ \\
         & & & & & & \\
         \hline
         & & & & & & \\
         & & 90 & -1.2 & 3.5 & (3$^{+2}_{-2}$)$\times$10$^{14}$ & & \\
         & & & & & & \\
         & & 90 & 3.4 & 3.9 & (8$^{+2}_{-3}$)$\times$10$^{14}$ & & \\
         H$_2$CO$^{(c)(d)}$ & & & & & & \\
         & & 90 & 9.1 & 5.1 & (3$^{+3}_{-2}$)$\times$10$^{14}$ & & \\
         & & & & & & \\
         & & 13$^{(e)}$ & 3.6$\pm$0.4 & 0.40$\pm$0.01 & $<$~6$\times$10$^{13}$ & & \\
         & & & & & & \\
         \hline
         & & & & & & \\
         & 0.30 & 90 & 2.1$\pm$0.3 & 2.8$\pm$0.6 & (2.6$^{+0.3}_{-0.2}$)$\times$10$^{12}$ & & \\
         D$_2$CO$^{(b)(c)(f)}$ & & & & & & \\
         & 1.00 & 90 & 2.1$\pm$0.3 & 2.8$\pm$0.6 & (4.3$^{+0.4}_{-0.4}$)$\times$10$^{12}$ & & \\
         & & & & & & \\
         \hline
    \end{tabular}
    \label{table_gauss}
    \footnotesize{\\$^{(a)}$ Upper limit $N_{\rm TOT}$ for absorption in HDCO was obtained by fixing V$_{\rm LSR}$ and FWHM using H$_2$CO absorption component and noting maximum $T_{\rm ex}$.}
   \footnotesize{\\$^{(b)}$ The HDCO and D$_2$CO values have been corrected for beam dilution.}
    \footnotesize{\\$^{(c)}$ $T_{\rm ex}$ was fixed to 90 K to model the emission components of H$_2$CO and D$_2$CO using the $T_{\rm ex}$ obtained for HDCO.}
    \footnotesize{\\$^{(d)}$ The FWHM and V$_{\rm LSR}$ values for H$_2$CO were fixed using the values obtained for HDCO.}
    \footnotesize{\\$^{(e)}$ An excitation temperature of 13 K for the envelope was used to model the absorption feature in H$_2$CO.}
    \footnotesize{\\$^{(f)}$ The FWHM and V$_{\rm LSR}$ values for D$_2$CO were fixed using L-M single Gaussian fitting}
    \end{table*}
\noindent Notably from our results, the FWHM of the components seen in the formaldehyde spectra show similarities with those of methanol (\citealt{vastel}), again providing support for the theory that these have a common origin (refer to Sect. \ref{origin}). It should be noted, however, that the number of higher s/n CH$_3$OH transitions analysed means the FWHM values for these emission components are better constrained than for the formaldehyde emission presented in this work.
\\
\\
Due to the presence of two H atoms in H$_2$CO, either of which may be deuterated to form HDCO, we calculate the D/H ratios using the following equation: 
\begin{gather}
\rm{\frac{D}{H}=\frac{N(HDCO)}{N(H_2CO)}\times0.5},
\label{eq.D/H}
\end{gather}
where N(HDCO) and N(H$_2$CO) represent the column densities ($N_{\rm TOT}$) of HDCO and H$_2$CO, respectively. Using the N(HDCO) results shown in Table \ref{table_gauss}, we calculate D/H ratios of 0.02$^{+0.02}_{-0.01}$ for the component at approximately -1 km~s$^{-1}$, 0.015$^{+0.008}_{-0.004}$ for the component at approximately 3 km~s$^{-1}$ and 0.01$^{+0.03}_{-0.01}$ for the component at approximately 9 km~s$^{-1}$ (the latter computed using only the unblended transition). This gives an average D/H ratio for HDCO in this source of 0.02$^{+0.02}_{-0.01}$ calculated using Eq. \ref{eq.D/H}. However, for D$_2$CO, we cannot conclude with any certainty where the emission is originating from or which of the multiple components detected in H$_2$CO and HDCO our single D$_2$CO emission component corresponds to. Therefore, while we can use our results to obtain upper limit values for N$_{\rm TOT}$, we cannot constrain the D/H ratio for D$_2$CO with any kind of certainty. 
\\
\\
For comparison, \citet{persson} found a D/H ratio of 0.03$\pm$0.01 in HDCO 
towards the protobinary IRAS16293-2422 B, within uncertainty of the value found in the present work towards [BHB2007] 11. It should be noted, however, that IRAS 16293-2422 B harbours a hot corino as interstellar Complex Organic Molecules (iCOMs) have been observed towards the centre of the hot core (\citealt{bottinelli}, \citealt{pineda12}, \citealt{jorgensen}). It has been shown that emission of methanol, the simplest iCOM, originating from the protostars of [BHB2007] 11 is optically thick and is therefore not centred on the hot core as one would expect for a hot corino (\citealt{vastel}). Meanwhile, \citet{bianchi} found D/H for HDCO towards the Class I source SVS13-A to be 0.086, which is also higher than the average value obtained for [BHB2007] 11. 
Similarly, \citet{zahorecz} obtained D/H ratios of 0.05 in HDCO towards a sample of high-mass star-forming regions, which is, again, higher than the average value obtained for [BHB2007] 11. 
\\
\\
In summary, the D/H ratio in HDCO obtained in this work is consistently low compared with values obtained from other deuteration studies, although the value seen is within error of that found towards the most similar source that we compare with: another protobinary, namely, IRAS 16293 2422 B
. Despite this, it is worth noting that our D/H calculation assumes that the H$_2$CO emission is uniform across our extraction region where HDCO 
is emitting. However, as previously discussed, the H$_2$CO emission shows similarity to that of CH$_3$OH, which would not show a uniform spectrum as compact emission regions have been detected (see \citealt{vastel}). This means that compact H$_2$CO emission regions are also likely to exist, however our current resolution is insufficient to confirm this through direct observation. Therefore, to account for this, the D/H ratios for [BHB2007] 11 presented in this work are to be taken as upper limit values. 
\section{Conclusions}\label{Conclusions}
We present detections of formaldehyde and its deuterated isotopologues towards the protobinary system [BHB2007] 11, observed using the ALMA interferometer within the context of the FAUST Large Program. From this data, we present column density measurements of all species obtained using radiative transfer analysis along with D/H ratios calculated using both HDCO and D$_2$CO for the first time. We also detect for the first time a large scale velocity feature separate from that previously observed in CO 2-1 and, following the results of the application of a kinematic model, tentatively propose the detection of a second large scale asymmetric molecular outflow launched by a wide-angle disk wind. The conclusions of this work can be summarised as follows:
\begin{itemize}
    \item We detect one H$_2$CO transition - 3(0,3)-2(0,2) at 218.222192 GHz - two HDCO transitions - 4(1,4)-3(1,3) at 246.9246 GHz and 4(2,2)-3(2,1) at 259.03491 GHz - and one D$_2$CO transition - 4(0,4)-3(0,3) at 231.410234 GHz - in this sample. The undeuterated H$_2$CO isotopologue seems to be ubiquitous, with clear presence in both the disk and extended envelope of the source. There is also some absorption associated with this transition due to the cold envelope oriented along the line of sight. The emission also traces extended velocity components oriented along the NW-SE axis perpendicular to the previously detected outflow feature. The morphology of the deuterated isotopologue transitions are consistent with each other and show much more compact emission only originating from within the circumstellar disk.
    \item Radiative transfer analysis has shown that there are three velocity components associated with both deuterated and undeuterated formaldehyde in this source, consistent with other FAUST observations. The velocity values of these components are consistent with the conclusion that the two components showing strong redshift and blueshift are originating from the southern Source B and the low velocity component is originating from the northern Source A, again consistent with other organic species. This is also consistent with the theory that the Source B is associated with high velocity small scale streamer material and a preferential accretion given that it is believed to be the less massive of the two protostars.
    \item Radiative transfer analysis has also enabled the column densities of H$_2$CO - (3-8)$\times$10$^{14}$ cm$^{-2}$ - as well as both HDCO isotopologues - (0.8-2.9)$\times$10$^{13}$ cm$^{-2}$ -  and the single D$_2$CO isotopologue - (2.6-4.3)$\times$10$^{12}$ cm$^{-2}$ -  
    to be obtained, enabling the measurement of the D/H ratio towards this source for the first time; this were found to be 0.02$^{+0.02}_{-0.01}$ averaged across the multiple HDCO velocity components
    . This D/H ratio in HDCO is consistent when compared to what has been previously observed in a similar protobinary object.
    \item Following analysis using a kinematic model (refer to the Appendix for full analysis), we propose that the extended features traced by the H$_2$CO emission are associated with a second asymmetric molecular outflow in addition to the one previously detected by \citet{Alves17}. Other molecular tracers of this feature need to be observed and analysed to distinguish between these two scenarios, with a study underway utilising FAUST observations of multiple dynamic tracers.
\end{itemize}
\begin{acknowledgements}
This project has received funding from (i) the European Union’s Horizon 2020 research and innovation programme under the Marie Skłodowska-Curie grant agreement No 811312 for the project "Astro-Chemical Origins” (ACO) and (ii) the European Research Council (ERC) under the European Union’s Horizon 2020 research and innovation programme, for the Project "The Dawn of Organic Chemistry" (DOC), grant agreement No 741002. This paper makes use of the following ALMA data: ADS/JAO.ALMA2018.1.01205.L. ALMA is a partnership of ESO (representing its member states), NSF (USA) and NINS (Japan), together with NRC (Canada), MOST and ASIAA (Taiwan), and KASI (Republic of Korea), in cooperation with the Republic of Chile. The Joint ALMA Observatory is operated by ESO, AUI/NRAO and NAOJ. I.J.-S. acknowledges support from grant No. PID2019-105552RB-C41 by the Spanish Ministry of Science and Innovation/State Agency of Research MCIN/AEI/10.13039/501100011033 and by “ERDF A way of making Europe." M.B. acknowledges the support from the European Research Council (ERC) Advanced Grant MOPPEX 833460.
\end{acknowledgements}
\bibliographystyle{aa}
\bibliography{bhb07-11_paper}
\newpage
\appendix
\section{Kinematic Streamer Modelling}
\begin{figure}[!h]
\centering
\begin{subfigure}{}
\includegraphics[width=250pt]{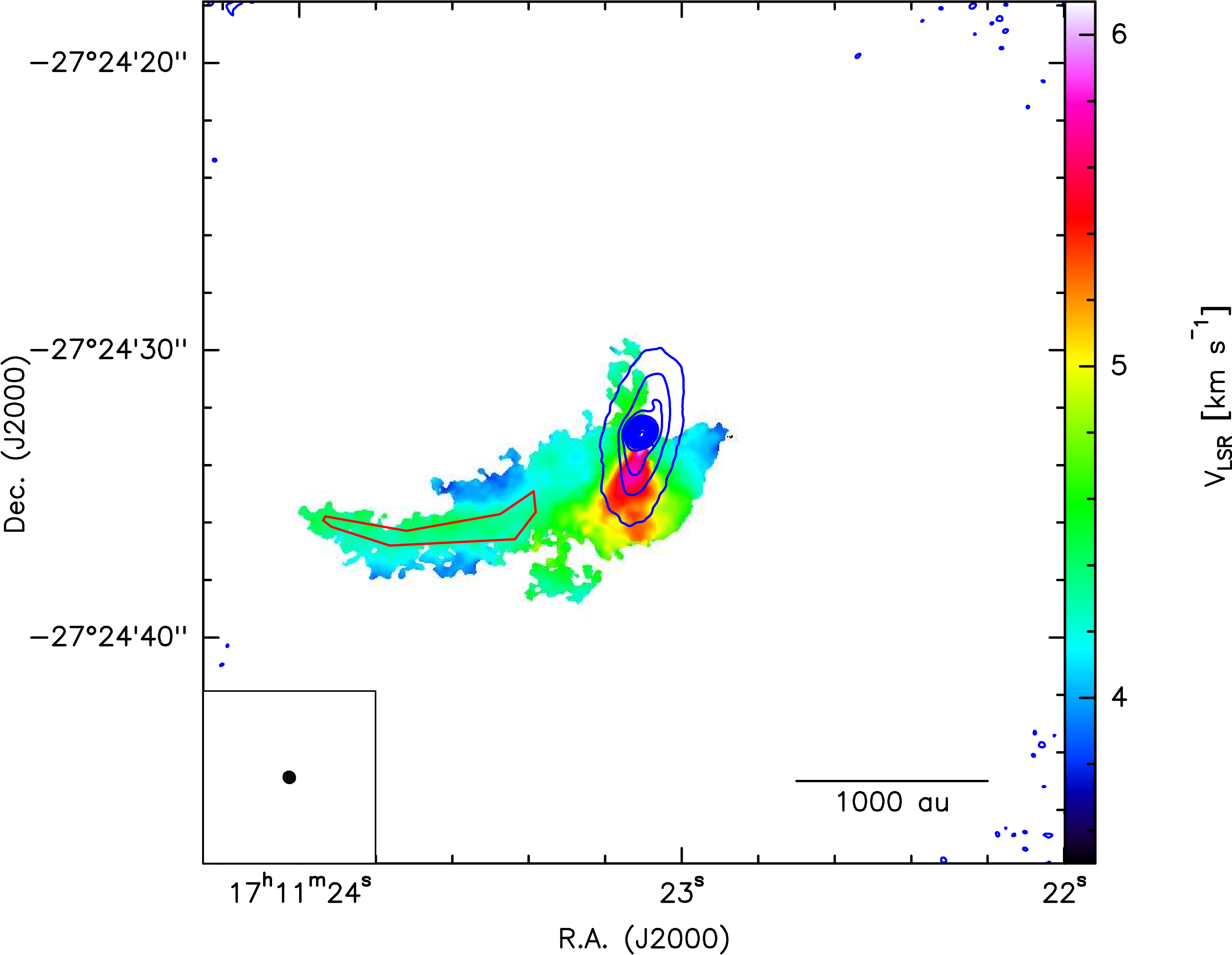}
\end{subfigure}
\begin{subfigure}{}
\includegraphics[width=250pt]{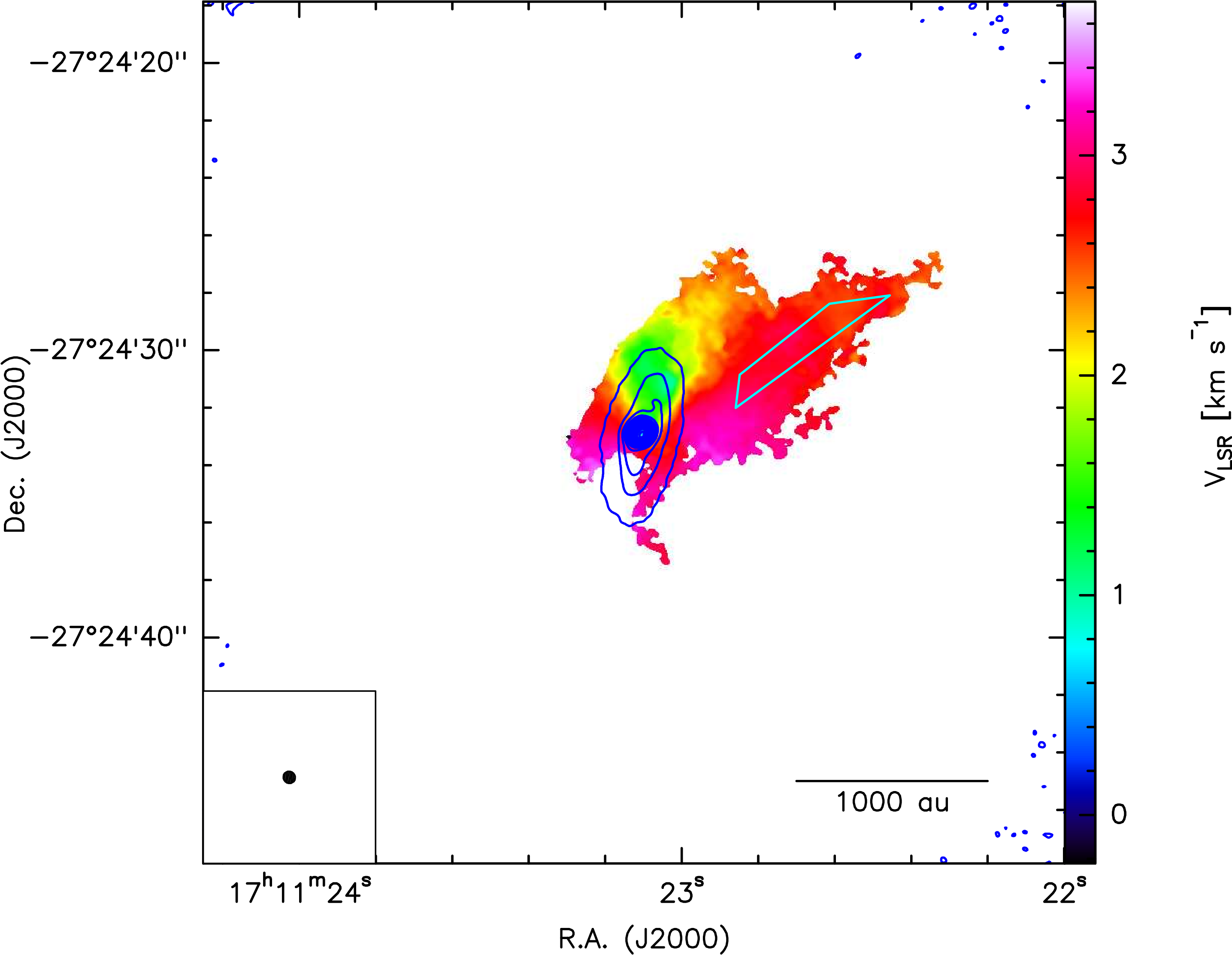}
\end{subfigure}
\caption{Centroid velocity maps from pixel-by-pixel Gaussian analysis showing pixels with intensity above 5$\sigma$ for the redshifted (upper panel; integrated between 3.5 and 6 km~s$^{-1}$) and blueshifted (lower panel; integrated between 0 and 3.5 km~s$^{-1}$) features overlaid with continuum emission (blue contours) between 4$\sigma$ and 300$\sigma$ in 50$\sigma$ intervals ($\sigma$=0.19 mJy~beam$^{-1}$). Polygons representing the pixels included in the kinematic model of the streamers are shown in red (upper panel) and cyan (lower panel). The ALMA synthesised beam is shown in the lower left corner.}
\label{fig.5sigma}
\end{figure}

\begin{figure}[!h]
    \centering
    \begin{subfigure}{}
    \includegraphics[width=250pt]{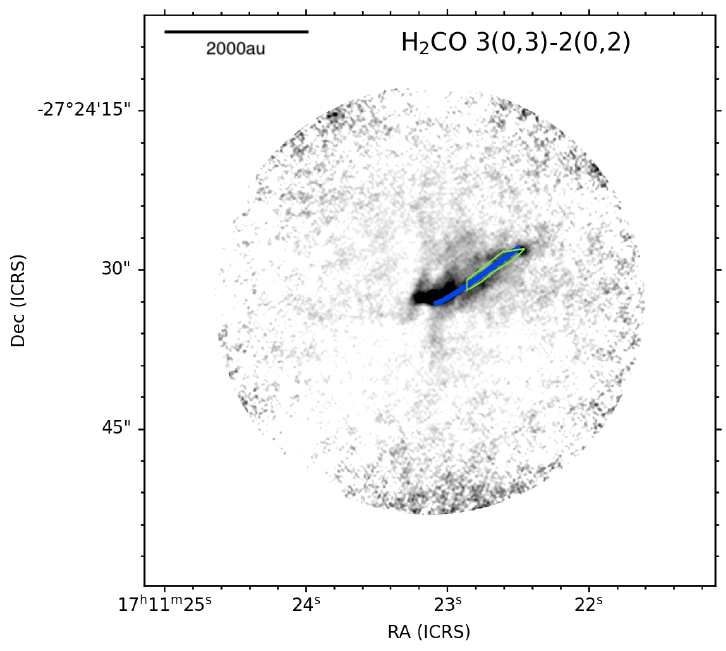}
    \end{subfigure}
    \begin{subfigure}{}
    \includegraphics[width=250pt]{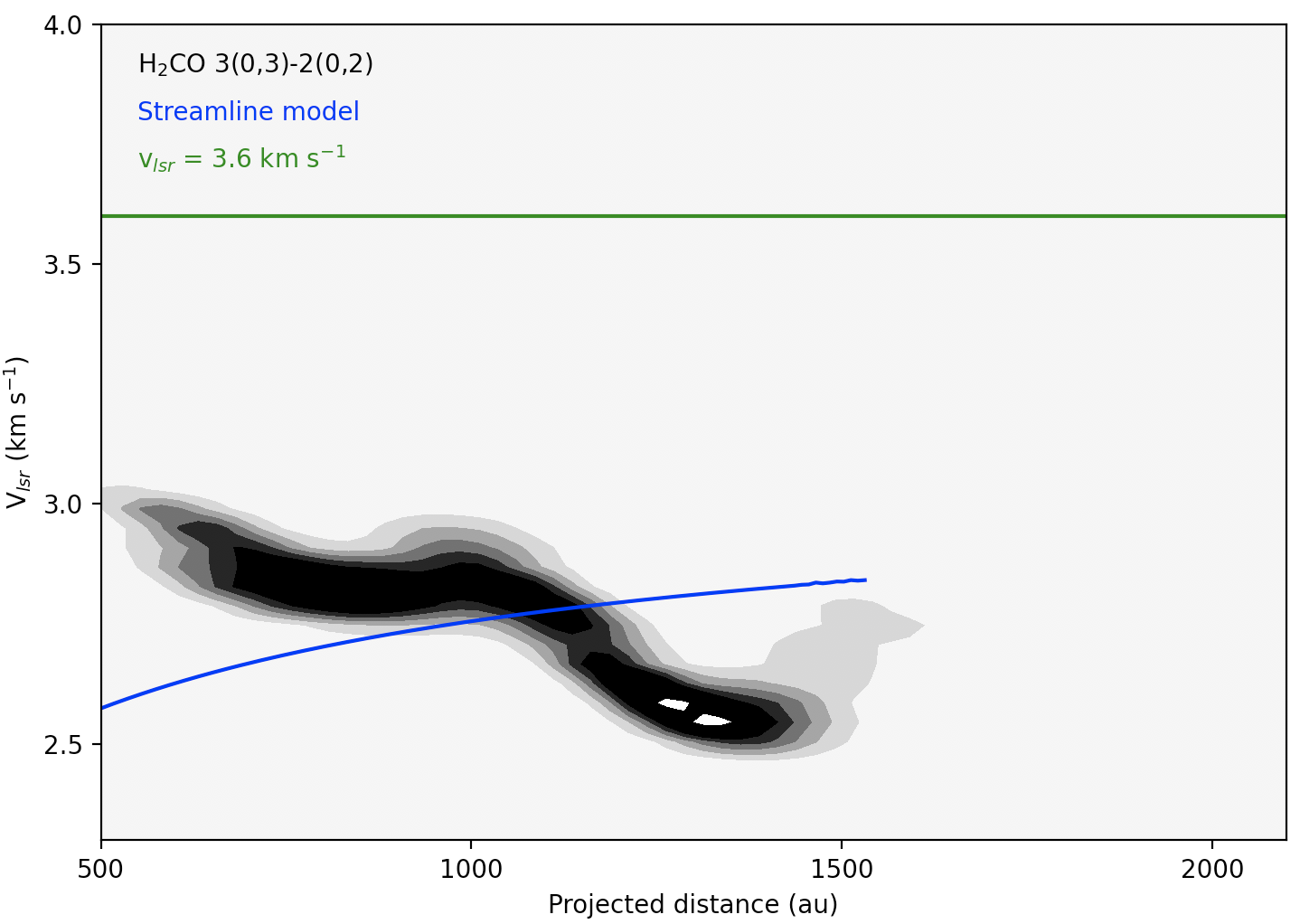}
    \end{subfigure}
    \caption{Streamer modelling results for blueshifted streamer. Upper panel: streamline (blue) generated using best fit initial parameters superimposed onto the H$_2$CO 3(0,3)-2(0,2) moment 0 map first presented in the left panel of Fig. \ref{fig.H2CO_cont}. Lower panel: plot showing velocity of streamer as a function of distance from source centre for raw data (black) and model (blue), with V$\rm_{LSR}$ value shown in green.}
    \label{fig.bluestream}
\end{figure}

\begin{figure}[!h]
    \centering
    \begin{subfigure}{}
   \includegraphics[width=250pt]{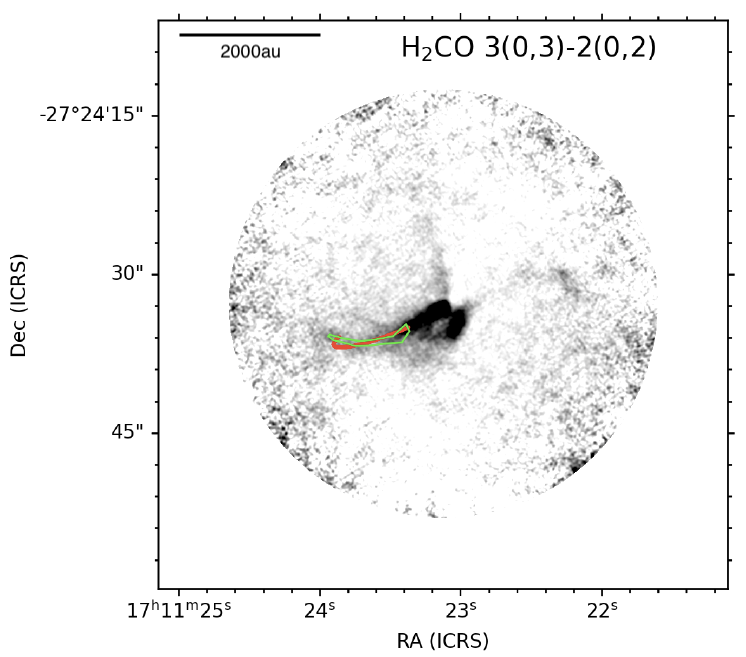}
   \end{subfigure}
   \begin{subfigure}{}
   \includegraphics[width=250pt]{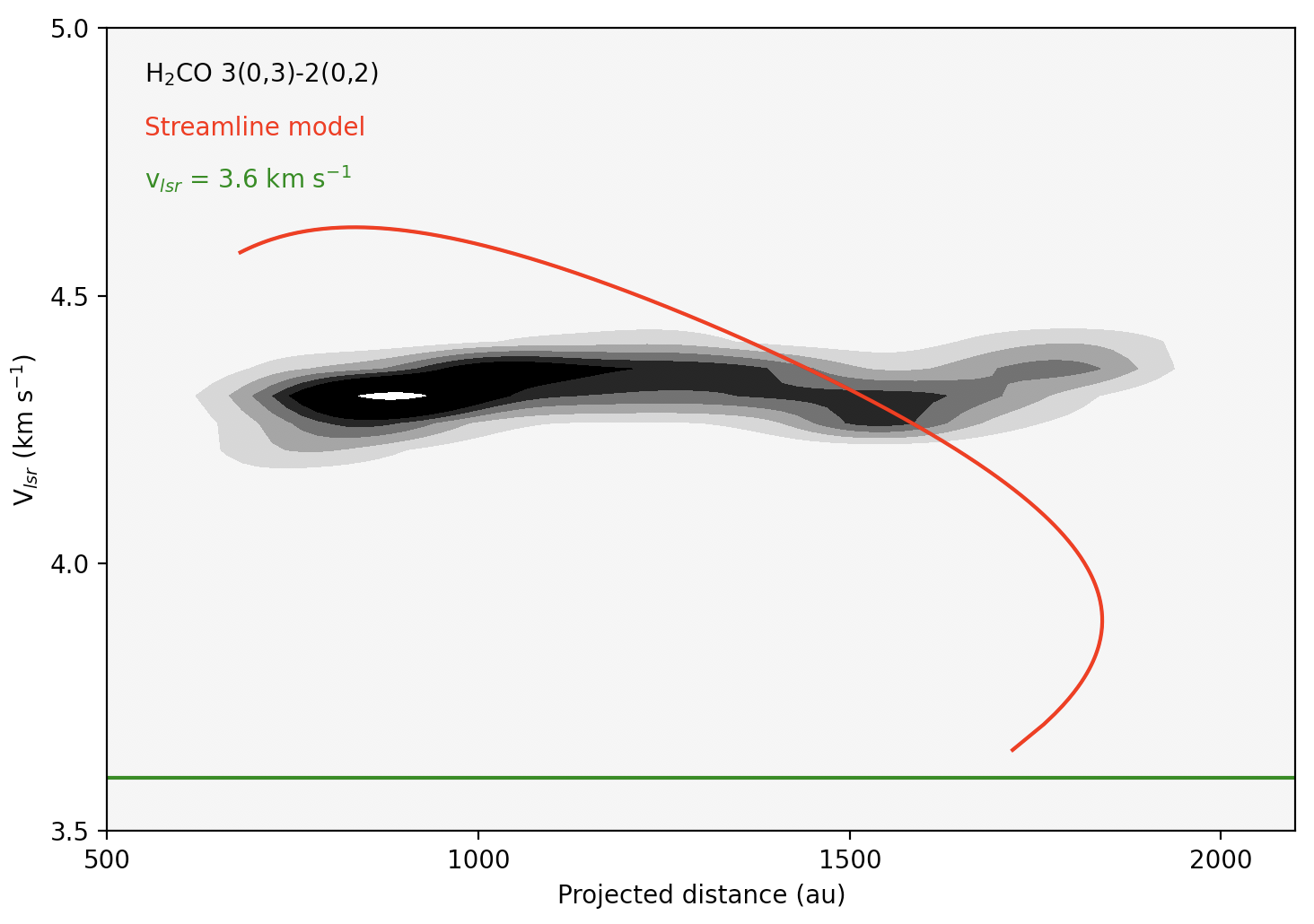}
   \end{subfigure}
   \caption{Streamer modelling results for redshifted streamer. Upper panel: streamline (red) generated using best fit initial parameters superimposed onto moment 0 map. Lower panel: plot showing velocity of streamer as a function of distance from source centre for raw data (black) and model (red), with V$\rm_{LSR}$ value shown in green.}
  \label{fig.redstream}
\end{figure}

As mentioned in Sect. \ref{H2CO}, our H$_2$CO emission clearly traces large scale dynamic material extending along the NW-SE axis which does not show alignment with any other feature previously detected in this source. It is already known that this source is dynamic as both small scale streamers within the circumbinary disk (\citealt{alves_19}) and a large scale bipolar molecular outflow (\citealt{Alves17}) have been seen; indeed, it is possible that interaction between the small scale streamers and quiescent gas in the circumstellar region is producing a shocked region, causing the non-thermal sublimation of organic material (e.g. CH$_3$OH and H$_2$CO) from the dust grain surface (\citealt{vastel}).
\\
\\
We implemented a streamline-generating model (\citealt{pineda}) in order to probe this dynamic feature with the knowledge that, if the feature is indeed streamer-like in nature, kinematic information such as the free-fall timescale and infall rate could be obtained as well as the extent and velocity relation of the feature (as previously obtained for the small-scale streamers by \citet{alves_19} using a similar method). However, due to a lack of a clear acceleration or deceleration, as well as a velocity relation that is flat or even outflowing, this investigation instead led to the conclusion that the feature is most likely not streamer-like in nature and, in fact, is most likely a second molecular outflow in addition to the previously detected feature seen in CO 2-1 by \citet{Alves17}.
\\
\\
In order to investigate the kinematics of the dynamic features detected for the first time in this source, we use the same model that was previously used to investigate the streamer detected in the protobinary system Per-emb-2 in \citet{pineda}. In this source, the streamer is chemically fresh material composed of carbon chain species originating from outside the dense core (more than 10,500 au) and infalling down to the disk scale. We focus on the region in [BHB2007] 11 with bright detections at large scales (approximately 2000 au), therefore we require a minimum intensity threshold of 5$\sigma$ for a pixel to be included in our modelling. To this end, prior to the modelling, we ran a pixel-by-pixel Gaussian fitting analysis using \textit{pyspeckit} on the redshifted and blueshifted features (individually) in order to identify these pixels for the model; the resultant maps can be seen in Fig. \ref{fig.5sigma}.
\noindent As can be seen from these pixel maps, we are clearly tracing two distinct velocity features in both the redshifted and blueshifted regime. The location and alignment of the high-velocity component suggest that this is originating from the disk. As confirmed in Sect. \ref{H2CO}, this coincides with the high-velocity component of H$_2$CO emission (see also Fig. \ref{fig.streamcont}). Using this, we define polygons that exclude the high-velocity disk emission and focus on the region with a clear and cohesive dynamic component. These are the polygons that are used in the modelling. 
\\
\\
The model generates streamlines assuming that material is free-falling (with conserved angular momentum) to a central object and uses analytic solutions from \citet{mendoza} to generate Cartesian co-ordinates for the position and velocity of a parcel of mass within a streamline, where the \textit{xy} plane is defined as the disk plane and the \textit{z} axis is along the angular momentum vector. Along with the central mass of the system (taken from \citealt{vastel}), the input parameters for the model are the spherical co-ordinates of the initial position and radial velocity of this parcel of mass within a cloud: $r_0$ (initial distance), v$_{\rm r,0}$ (radial velocity), $\theta_0$ (position angle with respect to the \textit{z} axis) and $\phi_0$ (initial angle within the disk plane), along with $\Omega_0$ (initial angular velocity of the cloud itself). In order to obtain the position and velocity values with respect to the observer, two rotations are applied due to the inclination angle \textit{i} and the disk position angle (both taken from \citealt{Alves17}). 
$r_0$ is estimated by comparing the extent of the generated streamline to that seen in the integrated intensity maps shown in the upper panels of Figs. \ref{fig.bluestream} and \ref{fig.redstream}, $\phi_0$ and $\theta_0$ are estimated by comparing the trajectory of the generated streamlines to the data both in the plane of the sky (again, using the integrated intensity maps shown in the upper panels of Figs. \ref{fig.bluestream} and \ref{fig.redstream}), as well as through comparison between the velocity-distance relation of the generated streamline to the raw data (see the lower panels of Figs. \ref{fig.bluestream} and \ref{fig.redstream}, while 
v$_{\rm r,0}$ is estimated by matching the propagation of velocity values along the length of the streamer, again using the velocity-distance maps shown in the lower panels of Figs. \ref{fig.bluestream} and \ref{fig.redstream}. 
\\
\\
Figs. \ref{fig.bluestream} and \ref{fig.redstream} show the results of the streamer modelling using these polygons with the initial parameters optimised to the data shown in Fig. \ref{fig.5sigma}. The upper panel of each figure shows the modelled streamer superimposed onto the moment 0 map for H$_2$CO 3(0,3)-2(0,2) while the lower panel shows the streamline velocity as a function of distance from the source. Note that the modelled trajectory on the plane of the sky is reasonable, however, the velocity relation is not. The velocity relation is very hard to reproduce and it is possible that either a problem with the assumption of conservation of angular momentum or the feature being located too close to the plane of the sky could be the reason behind the limited ability to constrain the model in this case. From these lower panels, it should also be acknowledged that the velocity relation seen from these features does not exhibit the inward acceleration expected from a streamer feature. For the redshifted streamer, the velocity relation is almost completely flat, which could be another consequence of the proximity of the feature to the plane-of-sky. For the blueshifted feature, we report a small outward velocity trend, which suggests an outflow rather than a streamer. However, as previously discussed, a bipolar outflow launched at symmetric positions with respect to the disk, approximately 110 au from the dust peak emission, has been traced by CO 2-1 in this source by \citet{Alves17}. This outflow is not spatially consistent with the features seen in H$_2$CO (refer to Fig. \ref{fig.streamcont}), while no evidence of the latter is seen in the outflow-tracing CO 2-1 observations. 
\noindent 
Assuming an infalling symmetric streamer, we would extract H$_2$CO spectra using the polygons shown in Figs. \ref{fig.bluestream} and \ref{fig.redstream} in order to obtain values for the H$_2$CO column density within each modelled streamer and use this to obtain a value for the column density of H$_2$ (N(H$_2$) in cm$^{-2}$) in each regime, from which we would use Eqs. \ref{eq.mass}, \ref{eq.freefall} and \ref{eq.accretion} to obtain the free-fall timescale ($t_{\rm ff}$) and infall rate ($\dot{M}_{\star}$), respectively:
\begin{gather}
M_{TOT}=\mu~m_H~d^2\delta~x\delta~y~N(H_2),
\label{eq.mass}
\end{gather}
\begin{gather}
t_{ff}=\sqrt{\frac{R^3}{GM}},
\label{eq.freefall}
\end{gather}
\begin{gather}
\dot{M_{\star}}=\frac{M_{TOT}}{t_{ff}},
\label{eq.accretion}
\end{gather}
where $\mu$ is the molecular weight (2.7, taking into account H$_2$, He and heavy element contributions), $m_{\rm H}$ is the atomic weight of hydrogen ($1.67\times10^{-24}$~g), $d$ is the distance to the star (163~pc - \citealt{dzib}), $\delta$~x$\delta$~y is the pixel size, \textit{M} represents the mass enclosed by radius \textit{R} (for which we use R = r$_0$ as found by the streamer model, \textit{G} is the gravitational constant and $L_{\rm bol}$ represents the bolometric luminosity of the source (taken as $1.68\times10^{27}$~W - \citealt{sandell}), all in cgs units. We could then compare our estimate for the accretion rate to the value of approximately 10$^{-5}~$M$_{\odot}$~year$^{-1}$ previously obtained for this source by \citet{alves_19} using a similar method. Therefore, material delivered via the potential infalling streamers might be comparable to the value delivered to the protostar. This is similar to what is seen in other objects (\citealt{pineda}, \citealt{valdiva}).
\\
\\
As a result, we propose that we are tracing a large scale asymmetric molecular outflow formed as the molecular gas is displaced from the cavity and evacuated by a disk wind. It is difficult to pinpoint the driving points. We see an asymmetric outflow detected in CO launched at the disk edge (approximately 100 au), as well as asymmetric flowing material traced by H$_2$CO at a much larger (envelope) scale. The lack of blueshifted H$_2$CO emission in the northeastern region and the lack of redshifted H$_2$CO emission in the southwestern region is puzzling but may be due to less gas being affected by the outflow due to lower densities and temperature. CO, which is tracing less dense material than H$_2$CO, may be hidden behind optically thick larger scale envelope emission which is being filtered out by the interferometer at velocities close to that of the systemic velocity. Recently, outflowing gas launched by an extended disk wind from a Keplerian disk with asymmetric, monopolar cavity walls has been detected (\citealt{bjerkali}), while another example of such a wide angle wind has been identified by \citet{torrelles}, although the latter towards a high-mass protostellar system. Therefore, wide-angle winds could be a natural outcome of the process of both low-mass and high-mass star formation. Further, archival ALMA observations tracing CO 3-2 towards this source (project number 2019.1.01566) show morphological evidence of the dynamic feature along with the expected cavity that would be associated with such a wide-angle disk wind. Therefore, while it is not possible with the present data alone to conclude on the nature of the dynamic material traced for the first time here, other molecular tracers detected towards these features would be very useful in contributing further evidence; indeed, an investigation to this end will be published in an upcoming study (Martinez-Henares et al. in prep.). In particular, this study will combine two transitions of CO from the archive with FAUST observations of multiple dynamic tracers with the primary goal of providing firm conclusions as to the nature of this newly detected dynamic feature of [BHB2007] 11.
\end{document}